\documentclass[reqno,12pt]{amsart}

\usepackage{amsmath,amssymb,amsthm,amsfonts,amsxtra}
\usepackage{fullpage}
\usepackage{graphicx}
\usepackage{caption}
\usepackage{subcaption}
\usepackage{enumitem}

\begin{document}

\newtheorem{thm}{Theorem}[section]
\newtheorem{cor}{Corollary}[section]
\newtheorem{lem}{Lemma}[section]
\newtheorem{prop}{Proposition}[section]

\numberwithin{equation}{section}

\def\Ref#1{Ref.~\cite{#1}}

\def\const{\text{const.}}
\def\Rnum{{\mathbb R}}
\def\sgn{{\rm sgn}}
\def\Dop{{\mathcal{D}}}
\def\nvec{{\mathbf n}}
\def\kvec{{\mathbf k}}
\def\sech{{\rm sech}}

\def\k{k}
\def\h{h_{\mathstrut}}

\def\X{{\mathrm X}}
\def\Esp{{\mathcal E}}
\def\var{\text{var.}}

\newlength{\mynoindent}
\settowidth{\mynoindent}{xxxxxxx}
\def\enumnoindent{\hspace{-\mynoindent}}

\tolerance=10000
\allowdisplaybreaks[3]

\title{Line-solitons, line-shocks, and conservation laws\\ of a universal KP-like equation
in 2+1 dimensions}

\author{
Stephen C. Anco$^1$
\lowercase{\scshape{and}}
M.L. Gandarias${}^2$
\lowercase{\scshape{and}}
Elena Recio$^{2}$ 
\\\\
${}^1$D\lowercase{\scshape{epartment}} \lowercase{\scshape{of}} M\lowercase{\scshape{athematics and}} S\lowercase{\scshape{tatistics}}\\
B\lowercase{\scshape{rock}} U\lowercase{\scshape{niversity}}\\
S\lowercase{\scshape{t.}} C\lowercase{\scshape{atharines}}, ON L2S3A1, C\lowercase{\scshape{anada}} \\\\
${}^2$D\lowercase{\scshape{epartment}} \lowercase{\scshape{of}} M\lowercase{\scshape{athematics}}\\
F\lowercase{\scshape{aculty of}} S\lowercase{\scshape{ciences}}, U\lowercase{\scshape{niversity of}} C\lowercase{\scshape{\'adiz}}\\
P\lowercase{\scshape{uerto}} R\lowercase{\scshape{eal}}, C\lowercase{\scshape{\'adiz}}, S\lowercase{\scshape{pain}}, 11510\\
}


\begin{abstract}
A universal KP-like equation in 2+1 dimensions,
which models general nonlinear wave phenomena exhibiting
$p$-power nonlinearity, dispersion, and small transversality,
is studied.
Special cases include the integrable KP (Kadomtsev-Petviashvili) equation and its modified version, as well as their $p$-power generalizations.
Two main results are obtained.
First, all low-order conservation laws are derived, 
including ones that arise for special powers $p$. 
The conservation laws comprise momenta, energy, and Galilean-type quantities,
as well as topological charges.
Their physical meaning and properties are discussed.
The topological charges are shown to give rise to integral constraints on initial data for the Cauchy problem. 
Second, 
all line-soliton solutions are obtained in an explicit form.
A parameterization is given using the speed and the direction angle of the line-soliton, and the allowed kinematic region is determined in terms of these parameters. 
Basic kinematical properties of the line-solitons are also discussed.
These properties differ significantly compared to those for KP line-solitons
and their $p$-power generalizations. 
A line-shock solution is shown to emerge when a special limiting case of the kinematic region is considered. 
\end{abstract}

\maketitle

\section{Introduction}

Numerous kinds of nonlinear wave phenomena 
exhibiting weak nonlinearity and dispersion for waves that have a small transverse component in 2+1 dimensions --- 
such as 
shallow water waves \cite{KadPet,AblSeg1979}, 
matter-wave pulses in Bose-Einstein condensates \cite{HuaMakVel}, 
ion-acoustic waves in plasmas \cite{InfRow}, 
and ferromagnets \cite{Leb} --- 
can be modelled by the Kadomtsev-Petviashvili (KP) equation \cite{KadPet}. 

Higher nonlinearities arise naturally in various nonlinear phenomena
\cite{PelSteKiv,KarBel,InfRow}, 
leading to a generalization of the KP equation with a $p$-power form
\cite{WanAblSeg,BouSau}
\begin{equation}\label{gKP}
(u_t +\alpha u^p u_x  +\beta u_{xxx})_x +\gamma u_{yy}=0,
\quad
p>0,
\end{equation}
called the gKP equation,
with constant coefficients $\alpha,\beta,\gamma$. 
The lowest-power nonlinearity $p=1$ produces the KP equation, 
and this is the only case in which it is known that the gKP equation is an integrable system.
It reduces to the $p$-power KdV equation when $u$ has no dependence on $y$. 

Recently,
a general modified KP-like equation has been derived in \Ref{RatBri}
by considering phase modulations of travelling waves in a universal nonlinear system in 2+1 dimensions. 
The resulting wave equation is given by 
\begin{equation}\label{genmKP}
(u_t +\alpha u^2 u_x +\epsilon uu_y +\kappa u_x\partial_x^{-1}u_y +\beta u_{xxx})_x +\gamma u_{yy}=0
\end{equation}
with constant coefficients $\alpha,\epsilon,\kappa,\beta,\gamma$.
This equation can be expected to model general nonlinear wave phenomena
exhibiting cubic nonlinearity, dispersion, and small transversality in 2+1 dimensions. 
(See \Ref{TsuOik,CheLiu,DasSar,VeeDan,RatBri,XuZhaLiFenTia} for physical applications).
In particular,
it is the most general KP-like equation that shares the same scaling symmetry group 
as the $p=2$ case of the gKP equation:
\begin{equation}
x\rightarrow \lambda x,
\quad
y\rightarrow \lambda^2 y,
\quad
t\rightarrow \lambda^3 t,
\quad
u\rightarrow \lambda^{-1} u 
\quad 
(\lambda\neq 0) . 
\end{equation}
Compared to the gKP equation for $p=2$, 
the modified equation \eqref{genmKP} contains two extra terms: $uu_y$ and $u_x\partial_x^{-1}u_y$,
which change sign under reflection $y \to -y$. 
When $u$ has no dependence on $y$, these terms vanish
and this equation reduces to the modified KdV equation. 

When the coefficient of the extra local term $uu_y$ is zero,
the resulting equation is known to be an integrable system if 
\begin{equation}\label{mKP}
\kappa^2 =-2\alpha\gamma/|\beta|, 
\quad
\sgn(\gamma/\beta)=-\sgn(\alpha/\beta)=1,
\quad
\epsilon=0, 
\end{equation}
which corresponds to the mKP equation \cite{KonDub1984}.  

Some basic aspects of the wave equations \eqref{gKP} and \eqref{genmKP}
are variational structures, line-soliton solutions, conservation laws, and symmetries.
These have been studied in \Ref{DaiHuaSunLiHu,BokZamFakKar,AncGanRec2018}
for the gKP equation, 
and in \Ref{GesHolSaaSim,KonDub1992,NazAliNae,ZhaXuJiaZho,AncGanRec2019}
for the integrable mKP equation.
No work of this kind has yet been done on the general modified equation \eqref{genmKP}. 

In the present paper,
we consider a $p$-power generalization of the general modified KP-like equation \eqref{genmKP}, 
given by 
\begin{equation}\label{mgKP}
(u_t +\alpha u^{p} u_x +\epsilon u^{\frac{p}{2}} u_y +\kappa u^{\frac{p}{2}-1}u_x\partial_x^{-1}u_y 
+\beta u_{xxx})_x + \gamma u_{yy}=0, 
\quad
p> 0 
\end{equation}
which we call the \emph{modified gKP equation}. 
Like the gKP equation and the general modified KP equation, 
it has a scaling symmetry and it reduces to the latter equation \eqref{genmKP} when $p=2$.
It will have applications in modelling wave phenomena that are characterized by 
higher nonlinearity, dispersion, and small transversality in 2+1 dimensions.

Apart from physical applications,
there are two general motivations for studying such a $p$-power family. 
One motivation is that the interactions of line-solitons 
depend sensitively on the value of $p$,
and specific integrability features 
such as asymptotic preservation of the shape and the speed of the line-solitons in collisions 
can be expected to break down for higher powers, particularly $p>2$. 
Another motivation involves studying the stability of line-solitons
as well as well-posedness of the Cauchy problem. 
Stability typically requires the existence of conserved mass and energy integrals 
and holds for $p$ not exceeding a critical value determined by their scaling invariance.

Our main goals here will be to determine 
the line-soliton solutions and the low-order conservation laws 
of the modified gKP equation \eqref{mgKP} 
for all nonlinearity powers $p>0$.
In particular,
our analysis will identify any special powers $p$ and special coefficient values
$\alpha$, $\epsilon$, $\kappa$, $\beta$, $\gamma$
for which either extra conservation laws are admitted 
or special kinematical features occur for the line-solitons.

We will show that a line-shock solution
emerges in a limiting case with $\alpha/\beta<0$. 
Line-shocks decay to zero in one asymptotic spatial direction
while in the opposite direction they asymptotically approach a non-zero value.
Such solutions do not exist for the gKP equation and thus are new phenomena
produced by the extra two terms $u^{\frac{p}{2}} u_y$ and $u^{\frac{p}{2}-1}u_x\partial_x^{-1}u_y$.

We will also find that, due to these extra terms,
non-symmetrical bright/dark pairs of line-solitons arise when $p/2$ is odd, 
whereas only symmetrical bright/dark pairs are supported by the gKP equation.

Further goals will be to study
the kinematical features of the line-solitons and the line-shocks 
and how they depend on the power $p$ and the coefficients $\alpha$, $\epsilon$, $\kappa$, $\beta$, $\gamma$,
as well as to investigate the physical and analytical properties of the conservation laws.
Interestingly,
when $\alpha/\beta <0$,
line-solitons and line-shocks cannot propagate purely in the $x$-direction --- namely, they must have some transverse component of velocity --- 
in contrast to the KP-like case $\alpha/\beta >0$.
Moreover, they exhibit an asymmetry when the sign of their angle of propagation with respect to the $x$ axis is reversed.

For both types of line solutions, 
their speed and direction angle are found to be determined entirely by their height and width, 
in contrast to the situation for KP line-solitons. 
Unlike typical approaches in the literature, 
we use a physical parameterization of the solutions,
which enables a better analysis of their properties. 

The kinematically allowed region in the parameter space of speed and angular direction
is determined by separating the analysis into four distinct cases 
given by the signs of $\alpha/\beta \gtrless 0$ and $\alpha\gamma\gtrless 0$. 
A significant qualitative difference in the resulting kinematic regions is found. 
For line-solitons, the size of the regions is independent of
the coefficients of the terms $u^{\frac{p}{2}} u_y$ and $u^{\frac{p}{2}-1}u_x\partial_x^{-1}u_y$ when $\alpha/\beta >0$, 
but has a sensitive dependence on these coefficients when $\alpha/\beta <0$. 
For line-shocks, the regions shrink to curves.
Additionally, when $\gamma/\beta >0$,
the speed is non-negative,
whereas when $\gamma/\beta <0$,
the speed can have either sign. 

The admitted conservation laws, for arbitrary $p$, are found to consist of
the $L^2$ norm and the mass,
as well as an energy and a $y$-momentum in the case
when the equation has a local Lagrangian structure.
There is rich structure of additional conservation laws when $p=1$ and $p=2$.
In the case $p=1$,
a Galilean momentum and a Galilean energy are admitted.
In the case $p=2$, 
both an energy 
and a linear combination of the $x$-moment of mass and the $y$-moment of momentum
are admitted when the equation has no local Lagrangian structure; 
two additional Galilean-like momentum quantities are admitted
in the Lagrangian case.
The sign properties of the energy will be determined,
and the critical powers for scaling invariance of the energy and the $L^2$ norm will be found. 

The modified gKP equation also possesses 
spatial flux conservation laws in all of the cases just mentioned. 
Their global form describes vanishing topological charges
given by a line-integral around any closed curve in the $(x,y)$-plane.
Some of these topological charges have an interesting relationship to
the $L^2$ norm, energy, and $y$-momentum, which will be discussed. 
In particular, 
integral constraints are shown to arise on initial data for the Cauchy problem, 
generalizing the well-known mass constraint \cite{MolSauTzv} on initial data 
for the KP equation. 

All of these results are new. 
The rest of the paper is organized as follows. 

First, in section~\ref{sec:potential},
the modified gKP equation \eqref{mgKP} is formulated as a local PDE 
by use of the potential $w$ given by $u=w_x$. 
The conditions on the coefficients for existence of a local Lagrangian structure
will be determined 
and contrasted with the local Lagrangian known \cite{AncGanRec2018}
for the gKP equation.  

Then, in section~\ref{sec:conslaws}, 
all low-order conservation laws of the modified gKP equation in potential form
are derived by the multiplier method. 
Computational aspects are summarized in an appendix. 
The physical meaning of each conservation law is described,
and the integral constraints are derived from those conservation laws that yield topological charges, 
following a method introduced recently in \Ref{AncRec2020}. 

Next, in section~\ref{sec:solns}, 
the line solutions $u=U(x+\mu y-\nu t)$ of the modified gKP equation 
are derived for $p>0$, 
where the parameters $\mu$ and $\nu$ determine the direction and the speed of the line wave.
The derivation uses the conservation laws of the equation to obtain 
a direct symmetry reduction to a separable ODE, 
by applying a new multi-reduction method developed in \Ref{AncGan2020}. 

In section~\ref{sec:kinematics},
the main kinematical properties of the line-solitons and line-shocks are discussed,
by considering a physical parameterization given by the speed and angular direction of these solutions. 

Finally, a few concluding remarks and goals for future work
are made in section~\ref{sec:remarks}.

\section{Potential form}\label{sec:potential}

The modified gKP equation \eqref{mgKP} is equivalent to a local PDE system 
\begin{equation}\label{mgKP-sys}
u_t +\alpha u^{2q} u_x +\epsilon u^{q}u_y + \kappa u^{q-1}u_x v +\beta u_{xxx} +\gamma v_y=0,
\quad
v_x = u_y
\end{equation}
where we have renamed the nonlinearity power 
\begin{equation}\label{mgKP-q}
q=\tfrac{1}{2}p >0 
\end{equation}
for convenience in the subsequent analysis
with $p$ fixed to be a positive integer.
Consequently, $q$ will be either a positive integer or a positive half-integer. 

This system \eqref{mgKP-sys} can be expressed as a single PDE 
by the introduction of a potential $w$ given by 
\begin{equation}\label{pot}
u=w_x,
\quad
v=w_y, 
\end{equation}
yielding
\begin{equation}\label{mgKP-pot}
w_{tx} + \alpha w_x^{2q} w_{xx} +\epsilon w_x^{q} w_{xy} +\kappa w_x^{q-1} w_{xx} w_y +\beta w_{xxxx} +\gamma w_{yy} =0,
\quad
q>0 .
\end{equation}
The potential $w$ has gauge freedom $w\to w + \chi(t,y)$
given by an arbitrary function $\chi(t,y)$.
Using this freedom, we can formally express $w$ in terms of $u$ by 
\begin{equation}\label{w-u-rel}
w(t,x,y) = \partial_x^{-1} u(t,x,y)
= \tfrac{1}{2}\Big( \int^x_{x_1} u(t,\zeta,y)\,d\zeta - \int^{x_2}_x u(t,\zeta,y)\,d\zeta \Big)
\end{equation}
where $x_1$ and $x_2$ can be chosen such that $w$ satisfies
a specified asymptotic condition as $x\to \pm\infty$.
For example,
$x_2=x_1=\pm\infty$ implies $w\to 0$ as $x\to\pm\infty$;
$x_2=-x_1=\infty$ implies $w\to \pm\int^{\infty}_{-\infty} u(t,x,y)\,dx$ as $x\to\pm\infty$. 

The modified gKP equation in potential form \eqref{mgKP-pot}
possesses the scaling symmetry 
\begin{equation}\label{mgKP-pot-scal}
x\rightarrow \lambda x,\
y\rightarrow \lambda^2 y,\
t\rightarrow \lambda^3 t,\
w\rightarrow \lambda^{1-\frac{1}{q}} w 
\quad
(\lambda\neq0). 
\end{equation}
By applying a general scaling transformation
\begin{equation}
  t\to \lambda_1 t,\
  x\to \lambda_2 x,\
  y\to \lambda_3 y,\
  w\to \lambda_4 w,
\end{equation}
where $\lambda_1,\lambda_2,\lambda_3,\lambda_4\neq0$,
we can fix three of the five coefficients $\alpha,\epsilon,\kappa,\beta,\gamma$
in equation \eqref{mgKP-pot}.
Specifically,
the coefficients transform as 
\begin{equation}
\alpha\to \lambda_1\lambda_4^{2q}\lambda_2^{-(2q+1)}\alpha,\
\epsilon\to \lambda_1\lambda_4^{q}\lambda_2^{-q}\lambda_3^{-1}\epsilon ,\
\kappa\to \lambda_1\lambda_4^{q}\lambda_2^{-q}\lambda_3^{-1}\kappa ,\
\beta\to \lambda_1\lambda_2^{-3}\beta,\
\gamma\to \lambda_1\lambda_2\lambda_3^{-2}\gamma,
\end{equation}
whence we can put
\begin{equation}\label{scaling}
|\alpha| =|\gamma|=\beta  = 1
\end{equation}
without loss of generality.

Note that in the case $\epsilon=\kappa=0$
we could further put $\sgn(\alpha)=1$ when $q$ is a half-integer (namely, $p$ is an odd integer).
But this is not possible if $\epsilon\neq 0$ or $\kappa\neq 0$ because
in this case the terms $w_x^{q}$  or $w_x^{q-1}$ would contain square roots
that require $w\geq 0$ with a subsequent restriction $\lambda_4>0$
in the scaling. 

Hereafter,
we will consider the modified gKP potential equation in the scaled form 
\begin{equation}\label{mgKP-pot-scaled}
w_{tx} + (\sigma_1 w_x^{2q} +a w_x^{q-1} w_y) w_{xx} + b w_x^{q} w_{xy}+ w_{xxxx} +\sigma_2 w_{yy} =0,
\quad
\sigma_1,\sigma_2=\pm 1,
\quad
q>0
\end{equation}
where $a,b$ are arbitrary constants such that $a\neq0$ or $b\neq0$, 
and where $q$ is either a positive integer or a positive half-integer. 
The corresponding form of the equation in terms of $u$ looks like
\begin{equation}\label{mgKP-scaled}
u_t + (\sigma_1 u^{2q} +a u^{q-1} \partial_x^{-1}u_y) u_x + b u^{q} u_y+ u_{xxx} +\sigma_2 \partial_x^{-1} u_{yy} =0, 
\end{equation}
which is an integrated form of equation \eqref{mgKP}. 
Note that the KP equation and the mKP equation are respectively given by (in scaled form) 
\begin{equation}\label{KP-case}
q=\tfrac{1}{2},
\quad
a=b=0, 
\quad
\sigma_1 =1
;
\end{equation}
\begin{equation}\label{mKP-case}
q=1,
\quad
a^2=2,
\quad
b=0 ,
\quad
\sigma_1=-1,
\quad
\sigma_2 =1
.
\end{equation} 
If we allow analytic continuations (specifically, $y\to iy$ and $u\to iu$),
then the conditions on $\sigma_1$ and $\sigma_2$ in mKP case
can be relaxed to $\sigma_1\sigma_2 =-1$. 

In the modified gKP equation \eqref{mgKP-scaled}, 
we will refer to $\sigma_1=1$ as the \emph{focussing} case, 
and $\sigma_1=-1$ as the \emph{defocussing} case,
in analogy with the mKdV equation. 
This distinction will be significant when line-soliton solutions are considered.

We will call $\sigma_2=1$ the \emph{normal dispersion} case 
and $\sigma_2=-1$ the \emph{sign-changing dispersion} case,
since for small amplitude solutions 
$w(x,t) \simeq A\exp(i(k_1x+k_2y-\omega t))$, with $|A|\ll 1$, 
the dispersion relation takes the form
$\omega = -k_1^3 +\sigma_2k_2^2/k_1$
which yields $\partial_{k_1}\omega = -(3k_1^2 +\sigma_2k_2^2/k_1^2)$ 
giving the group velocity in the $x$ direction. 
We see that, when $\sigma_2=1$, the group velocity has a single sign,
whereas when $\sigma_2=-1$, the sign of the group velocity changes
when $|k_2| = \sqrt{3}|k_1|^2$.

\subsection{Variational structure} 

A wave equation of the form $w_{tx} =F(w,w_x,w_y,\ldots)$ 
will be an Euler-Lagrange equation of a local Lagrangian in terms of $w$ 
iff the Helmholtz conditions \cite{Olv-book,Anc-review} are satisfied.
These conditions state that the Frechet derivative of the wave equation 
needs to be self-adjoint.
It is straightforward to show that Frechet derivative of the term $w_{tx}$ is self-adjoint,
and hence the existence of a local Lagrangian depends solely on whether 
the Frechet derivative of the term $F(w,w_x,w_y,\ldots)$ 
is self-adjoint. 

As shown in \Ref{AncGanRec2018}, 
the gKP equation has a local Lagrangian,
and this structure corresponds to a Hamiltonian formulation when the gKP equation is 
expressed as an evolution equation for $u$. 

The situation for the modified gKP equation, including the mKP equation, is quite different. 
The Frechet derivative of the potential form \eqref{mgKP-pot-scaled} of the equation 
is given by

\begin{equation}\label{frech}
\begin{aligned}
&
D_tD_x P
+ ( (\sigma_1 2q w_x^{2q-1} +a (q-1)w_x^{q-2} w_y)w_{xx} +bq w_x^{q-1} w_{xy} )D_x P 
\\&
+a w_x^{q-1} w_{xx} D_y P
+ ( \sigma_1 w_x^{2q} +a w_x^{q-1} w_y )D_x^2 P
+b w_x^{q} D_xD_y P 
+\sigma_2 D_y^2 P +D_x^4 P
\end{aligned}
\end{equation}
where $P=P(t,x,y)$. 
The adjoint Frechet derivative is obtained via multiplication by $Q=Q(t,x,y)$
followed by integration by parts, yielding $P$ times 
\begin{equation}\label{adjfrech}
\begin{aligned}
&
D_tD_x Q -D_x(( (\sigma_1 2q w_x^{2q-1} +a (q-1)w_x^{q-2} w_y)w_{xx} +bq w_x^{q-1} w_{xy} )Q)
\\&
-D_y( a w_x^{q-1} w_{xx} Q )
+ D_x^2(( \sigma_1 w_x^{2q} +a w_x^{q-1} w_y )Q)
+D_xD_y(b w_x^{q} Q)
+\sigma_2 D_y^2 Q +D_x^4 Q
\end{aligned}
\end{equation}
modulo total derivatives.
For the Frechet derivative to equal its adjoint,
expression \eqref{frech} minus expression \eqref{adjfrech} with $Q=P$
must vanish identically for all $P(t,x,y)$.
This yields the following necessary and sufficient condition for
existence of a local Lagrangian.

\begin{prop}\label{prop:L-H-structure}
The modified gKP equation \eqref{mgKP-pot-scaled} possesses a local Lagrangian in terms of the potential $w$ iff
\begin{equation}\label{mgKP-variational}
a=\tfrac{1}{2} bq . 
\end{equation}
The Lagrangian, $L$, is given by
\begin{equation}\label{mgKP-L}
L = 
-\tfrac{1}{2} w_tw_x
-\tfrac{1}{(2q+2)(2q+1)}\sigma_1 w_x^{2q+2}
-\tfrac{1}{2q+2} b w_x^{q+1} w_y
-\tfrac{1}{2}\sigma_2 w_y^2
+\tfrac{1}{2} w_{xx}^2 .
\end{equation}
\end{prop}

Note that the case $a=b=0$ corresponds to the gKP equation \eqref{gKP} (up to scaling) whose Lagrangian was obtain in \Ref{AncGanRec2018}.
Also note that the case $a\neq0$ and $b=0$ when the modified gKP equation \eqref{mgKP-pot-scaled}
does not possess a local Lagrangian structure in terms of $w$
includes the case of the mKP equation \eqref{mKP-case}. 

The implication of this result for existence of conservation laws
is fully discussed in the next section.

When the Lagrangian exists, there is a corresponding Hamiltonian structure 
\begin{equation}
u_t = D_x( \delta H/\delta u )
\end{equation}
where
\begin{equation}
H = \int_{\Rnum^2}\big( \tfrac{1}{2} u_x^2 -\tfrac{1}{2}\sigma_2 (\partial_x^{-1}u_y)^2 -\tfrac{1}{2(q+1)}b u^{q+1}\partial_x^{-1}u_y  -\tfrac{1}{2(q+1)(2q+1)}\sigma_1 u^{2q+2}\big) dxdy
\end{equation}
is the Hamiltonian functional,
and $D_x$ is a Hamiltonian operator.

\section{Conservation laws}\label{sec:conslaws}

Conservation laws are important in the analysis of nonlinear evolution equations 
by providing physical, conserved quantities as well as conserved norms needed
for studying well-posedness, stability, and global behaviour of solutions. 

For the modified gKP potential equation \eqref{mgKP-pot-scaled}, 
a local conservation law is a continuity equation
\begin{equation}\label{conslaw}
D_t T+D_x X+D_y Y=0
\end{equation}
holding for all solutions $w(x,y,t)$ of equation \eqref{mgKP-pot-scaled},
where $T$ is the conserved density, and $(X,Y)$ is the spatial flux,
which are functions of $t$, $x$, $y$, $w$, and derivatives of $w$. 
Note that $w_{tx}$ and all of its derivatives can be eliminated from $T,X,Y$ 
through expressing 
$w_{tx} = -\big( (\sigma_1 w_x^{2q} +a w_x^{q-1} w_y) w_{xx} + b w_x^{q} w_{xy}+ w_{xxxx} +\sigma_2 w_{yy} \big)$
from equation \eqref{mgKP-pot-scaled}. 

When solutions $w(x,y,t)$ are considered 
in a given spatial domain $\Omega\subseteq\Rnum^2$, 
every local conservation law yields a corresponding conserved integral 
\begin{equation}\label{conservedintegral}
\mathcal{C}[w]= \int_{\Omega} T\,dx\,dy
\end{equation}
satisfying the global balance equation
\begin{equation}\label{globalconslaw}
\frac{d}{dt}\mathcal{C}[w]= -\int_{\partial\Omega} (X,Y)\cdot\hat\nvec\,ds
\end{equation}
where $\hat\nvec$ is the unit outward normal vector of the domain boundary curve $\partial\Omega$, 
and where $ds$ is the arclength on this curve with clockwise orientation. 
This global equation \eqref{globalconslaw} has the physical meaning that
the rate of change of the quantity \eqref{conservedintegral} on the spatial domain 
is balanced by the net outward flux through the boundary of the domain. 

A conservation law is locally trivial \cite{Olv-book,BCA-book,Anc-review} 
if, for all solutions $w(x,y,t)$ in $\Omega$,
the conserved density $T$ reduces to a spatial divergence $D_x \Psi^x + D_y \Psi^y$ 
and the spatial flux $(X,Y)$ reduces to a time derivative $-D_t(\Psi^x,\Psi^y)$ 
modulo a spatial curl $(D_y\Theta,-D_x\Theta)$, 
since then the global balance equation \eqref{globalconslaw} becomes an identity. 
Likewise, two conservation laws are locally equivalent \cite{Olv-book,BCA-book,Anc-review} 
if they differ by a locally trivial conservation law, for all solutions $w(x,y,t)$ in $\Omega$. 
We will be interested only in locally non-trivial conservation laws. 

Because the modified gKP potential equation \eqref{mgKP-pot-scaled} in general
has no Lagrangian structure, 
Noether's theorem cannot be applied to derive conservation laws. 
Instead, its conservation laws arise from multipliers \cite{Olv-book,AncBlu2002b,BCA-book,Anc-review}
as follows. 

Any non-trivial conservation law \eqref{conslaw} 
can be expressed in an equivalent characteristic form \cite{Olv-book,BCA-book,Anc-review}
which is given by a divergence identity holding off of the space of solutions $w(x,y,t)$. 
For the modified gKP potential equation \eqref{mgKP-pot-scaled}, 
conservation laws have the characteristic form
\begin{equation}\label{gmkp-chareqn}
D_t\tilde T+D_x\tilde X+D_y\tilde Y=
( w_{tx} +(\sigma_1 w_x^{2q} +a w_x^{q-1} w_y) w_{xx} + b w_x^{q} w_{xy}+ w_{xxxx} +\sigma_2 w_{yy} )Q
\end{equation}
where $\tilde T$, $\tilde X$, $\tilde Y$, and $Q$ 
are functions of $t$, $x$, $y$, $w$, and derivatives of $w$,
and where the conserved density $\tilde T$ and the spatial flux $(\tilde X,\tilde Y)$
reduce to $T$ and $(X,Y)$ when restricted to all solutions $w(x,y,t)$ of equation \eqref{mgKP-pot-scaled}. 
This divergence identity is called the characteristic equation for the conservation law,
and the function $Q$ is called the conservation law multiplier. 
In general, $Q$ will be non-singular when it is evaluated on any solution $w(x,y,t)$. 
As a consequence, 
the characteristic equation of a conservation law is locally equivalent to the conservation law itself. 

It will be useful to note that pure leading derivatives of the modified gKP potential equation \eqref{mgKP-pot-scaled} consist of $w_{yy}$ or $w_{xxxx}$. 
If either leading derivative, and all of its differential consequences, 
have been eliminated from $T,X,Y$, 
then $Q$ will not contain those eliminated variables \cite{Olv-book,BCA-book,Anc-review}
and there will be a one-to-one correspondence between non-trivial conservation laws (up to equivalence) and non-zero multipliers. 

All multipliers $Q$ can be determined from the characteristic equation \eqref{gmkp-chareqn}
by use of the Euler operator \cite{Olv-book,BCA-book,Anc-review} 
$E_w$ with respect to $w$,
where this operator annihilates a function of $t$, $x$, $y$, $w$, and derivatives of $w$
iff the function is given by a total divergence. 
In particular, multipliers $Q$ are the solutions of the determining equation 
\begin{equation}\label{Q-deteqn}
E_w\big(( w_{tx} +(\sigma_1 w_x^{2q} +a w_x^{q-1} w_y) w_{xx} + b w_x^{q} w_{xy}+ w_{xxxx} +\sigma_2 w_{yy} )Q\big) =0
\end{equation}
holding off of solutions of equation \eqref{mgKP-pot-scaled}. 
This determining equation has a natural splitting with respect to either of 
the pure leading derivatives $w_{yy}$ or $w_{xxxx}$ and its differential consequences. 
The splitting yields a determining system that consists of the adjoint of the determining equation for symmetries \cite{Olv-book,BCA-book,Anc-review} 
plus additional determining equations analogous to Helmholtz-type equations 
\cite{Anc-review}. 
Consequently, multipliers have a characterization as adjoint-symmetries satisfying 
certain Helmholtz-type conditions \cite{BCA-book,Anc-review}. 

When $Q$ is specified to have any chosen form, 
with its differential order fixed with respect to $w$
and with no dependence on either of the pure leading derivatives $w_{yy}$ or $w_{xxxx}$
and its differential consequences, 
then the determining equation \eqref{Q-deteqn} can be solved in a similar way to the 
symmetry determining equation,
by splitting it with respect to all variables that do not appear in $Q$
so as to obtain an overdetermined system of equations on $Q$. 
Thus, multipliers can be found by similar computational steps used to find symmetries.

In the case \eqref{mgKP-variational} when the modified gKP potential equation \eqref{mgKP-pot-scaled}
has a variational structure,
every conservation law corresponds to a variational symmetry
$\X = \tau\partial_t + \xi^x\partial_x + \xi^y\partial_y + \eta\partial_u$ of the Lagrangian \eqref{mgKP-L}
through Noether's theorem.
Here the components $\tau$, $\xi^x$, $\xi^y$, $\eta$ of the symmetry generator
are functions of $t$, $x$, $y$, $w$, and derivatives of $w$. 
The Noether correspondence states that
the multiplier is given by the characteristic form of the symmetry:
\begin{equation}
Q =  \eta -\tau u_t - \xi^x u_x - \xi^y u_y . 
\end{equation}
In particular, the determining system for multipliers
can be shown to coincide with the determining system for variational symmetries.

For any given multiplier $Q$, 
the corresponding conserved density $\tilde T$ and spatial flux $(\tilde X,\tilde Y)$
can be obtained straightforwardly through a repeated integration process \cite{Wol,BCA-book,Anc-review}
applied to the terms in the righthand side of the characteristic equation \eqref{gmkp-chareqn}. 
This method can sometimes be lengthy or awkward, 
depending on the complexity of the righthand side expression.
A more direct method is to use a homotopy integral formula that inverts the Euler operator $E_w$. 
The simplest version of this formula appears in \Ref{BCA-book,Anc-review};
a more complicated general version (in the context of the variational bi-complex)
is given in \Ref{Olv-book}. 
Alternatively, 
since the modified gKP potential equation \eqref{mgKP-pot-scaled} possesses a scaling symmetry, 
there is an algebraic scaling formula \cite{Anc2003,BCA-book,Anc-review} 
which can be used to obtain an explicit expression for $\tilde T,\tilde X,\tilde Y$
whenever the corresponding conserved integral \eqref{conservedintegral} is not scaling invariant. 
However, 
both the scaling formula and the homotopy integral formula have the drawback that they 
do not directly yield the lowest possible differential order (up to equivalence)
for the conserved density $T$, 
whereas the integration by parts method can be applied in a way that does this. 

Typically, for wave equations, 
all multipliers that correspond to physical conservation laws 
such as energy and momentum 
are of a lower differential order than the given equation,
while multipliers of higher differential order are most often connected with integrability features of the given equation. 

Here we will explicitly find all low-order conservation laws of the modified gKP potential equation \eqref{mgKP-pot-scaled} for $p\neq 0$ 
by determining all multipliers with a differential order of less than four. 
This class of multipliers has the general form 
\begin{equation}\label{low-order-Q}
Q(t,x,y,w,\partial w,\partial^2 w,\partial^3 w) 
\end{equation}
where $\partial=(\partial_t,\partial_x,\partial_y)$. 
Note that any expression of this form \eqref{low-order-Q} is necessarily non-singular when it is evaluated on any $w(x,y,t)$ satisfying equation \eqref{mgKP-pot-scaled}. 
For the gKP case $a=b=0$, $q\neq 0$, 
all low-order conservation laws have been obtained in \Ref{AncGanRec2019}. 
Computational remarks are provided in the appendix.

\begin{prop}\label{prop:multrs}
All low-order multipliers \eqref{low-order-Q} admitted by 
the modified gKP potential equation \eqref{mgKP-pot-scaled} 
with $q\neq 0$, $a^2+b^2\neq0$, $\sigma_1^2=1$, $\sigma_2^2=1$ are given by
\begin{align}
& \label{multr1}
Q_{(1)} = w_x , 
\\
& \label{multr2}
Q_{(2)} = f(t) ,
\end{align}
where $f(t)$ is an arbitrary function. \\
All cases $q\neq 0$ and $a^2+b^2\neq0$ for which
the modified gKP potential equation \eqref{mgKP-pot-scaled} admits
additional low-order multipliers \eqref{low-order-Q} consist of: 
\begin{enumerate}[label=(\roman*)]
\item $a=\tfrac{1}{2}bq$
\begin{align}
& \label{multr3}
Q_{(3)} = w_t , 
\\
& \label{multr4}
Q_{(4)} = w_y ; 
\end{align}

\item $q=\tfrac{1}{2}$, $a=\tfrac{1}{4}b$
\begin{equation}\label{multr5}
Q_{(5)} = 3t w_t +x w_x +2y w_y +w ; 
\end{equation}

\item $q=\tfrac{1}{2}$, $a=0$
\begin{equation}\label{multr6}
Q_{(6)} = 
x -\sigma_1 t w_x ; 
\end{equation}

\item $q=1$, $a=\tfrac{1}{2}b$
\begin{equation}\label{multr7}
Q_{(7)} = 
bx-\tfrac{4}{3}\sigma_1 y w_x
+(\tfrac{8}{3}\sigma_1\sigma_2-b^2) t w_y ;
\end{equation}

\item $q=1$, $b^2=a^2+2\sigma_1\sigma_2$
\begin{equation}\label{multr8}
Q_{(8)} = 
(a+b)w_t
+\tfrac{2}{3}(2a-b)\sigma_1 w_x^3
+(a^2+ab-2\sigma_1\sigma_2)w_x w_y
+2(2a-b)w_{xxx} ; 
\end{equation}

\item $q=1$, $b=\tfrac{1}{2}(a+2\sigma_1\sigma_2/a)$
\begin{equation}\label{multr9}
Q_{(9)} = x -a\sigma_2 y w_x ; 
\end{equation}

\item $q=-2$, $a=-b$
\begin{equation}\label{multr10}
Q_{(10)} = y w_x -2\sigma_2 t w_y ; 
\end{equation}
\enumnoindent
and, with an arbitrary function $f(t)$, 
\item $a=bq$
\begin{equation}\label{multr11}
Q_{(11)} = 
f(t)y ; 
\end{equation}

\item $q=1$
\begin{equation}\label{multr12}
Q_{(12)} = 
f'(t)y + (a-b) w_x f(t) ; 
\end{equation}

\item $q=1$, $a=\tfrac{1}{2}b$, $b^2=-\tfrac{8}{3}\sigma_1\sigma_2$
\begin{equation}\label{multr13}
Q_{(13)} = 
w_y f(t) +(\tfrac{3}{16} \sigma_1\sigma_2 b x -\tfrac{1}{4}\sigma_2 w_x y) f'(t)
-\tfrac{3}{32}\sigma_1 b y^2 f''(t) ; 
\end{equation}

\item $q=1$, $b=0$, $a^2=-2\sigma_1\sigma_2$
\begin{align}
& \label{multr14}
Q_{(14)} = 
( y w_x +\tfrac{1}{2}\sigma_1 a x) f(t)
-\tfrac{1}{4} \sigma_1\sigma_2 a y^2 f'(t) , 
\\
& \label{multr15}
\begin{aligned}
Q_{(15)} = &
( \tfrac{3}{4} \sigma_1 w_t +\tfrac{3}{2} \sigma_1 a w_x w_y + w_x^3 +3\sigma_1 w_{xxx} )f(t)
-\tfrac{3}{4} a x w_x f'(t)
\\&\quad
+\tfrac{3}{8} \sigma_2( \sigma_1 y^2 w_x +ax y) f''(t) 
-\tfrac{1}{16} a y^3 f'''(t) .
\end{aligned}
\end{align}
\end{enumerate}
\end{prop}

These multipliers determine all non-trivial conservation laws of low order
admitted by the modified gKP potential equation \eqref{mgKP-pot-scaled}. 
A summary of the computation is provided in the appendix. 

\begin{thm}\label{thm:conslaws}
All low-order local conservation laws
admitted by the modified gKP potential equation \eqref{mgKP-pot-scaled} 
with $q\neq 0$ and $a^2+b^2\neq0$ 
are given by (up to equivalence)
\begin{subequations}\label{conslaw1}
\begin{flalign}
\quad&\begin{aligned}
T_{(1)} = & 
\tfrac{1}{2}w_{x}^2
, 
\end{aligned}&&
\\
\quad&\begin{aligned}
X_{(1)} = & 
w_x w_{xxx} 
-\tfrac{1}{2}w_{xx}^2 
+\tfrac{1}{2} (\sigma_1/(q+1)) w_x^{2(q+1)} 
+ (a/(q+1) w_x^{q+1}w_y 
-\tfrac{1}{2}\sigma_2 w_y^2
, 
\end{aligned}&&
\\
\quad&\begin{aligned}
Y_{(1)}= & 
\sigma_2 w_x w_y 
-(a-b-bq)/((q+1)(q+2)) w_x^{q+2}
; 
\end{aligned}&&
\end{flalign}
\end{subequations}

\begin{subequations}\label{conslaw2}
\begin{flalign}
\quad&\begin{aligned}
T_{(2)} = &
w_x f(t)
, 
\end{aligned}&&
\\
\quad&\begin{aligned}
X_{(2)} = & 
\big(
w_{xxx}
+(a/q) w_x^q w_y 
+(\sigma_1/(2q+1)) w_x^{2q+1} 
\big)  f(t)
- w f'(t)
, 
\end{aligned}&&
\\
\quad&\begin{aligned}
Y_{(2)}= & 
\big(
\sigma_2 w_y 
-(a-bq)/(q(q+1)) w_x^{q+1}
\big)  f(t)
; 
\end{aligned}&&
\end{flalign}
\end{subequations}

All cases $q\neq 0$ and $a^2+b^2\neq0$ for which
the modified gKP potential equation \eqref{mgKP-pot-scaled} admits
additional local conservation laws consist of:
\begin{enumerate}[label=(\roman*)]
\item $a=\tfrac{1}{2}bq$
\begin{subequations}\label{conslaw3}
\begin{flalign}
\quad&\begin{aligned}
T_{(3)} = & 
\tfrac{1}{2}w_{xx}^2
-\tfrac{1}{2}\sigma_2 w_{y}^2
-\tfrac{1}{2} (b/(q+1)) w_{x}^{q+1}w_y
-\tfrac{1}{2} (\sigma_1/((q+1)(2q+1))) w_{x}^{2(q+1)}
, 
\end{aligned}&&
\\
\quad&\begin{aligned}
X_{(3)} = & 
w_t w_{xxx} 
-w_{tx}w_{xx}
+\tfrac{1}{2}w_{t}^2 
+(\sigma_1/(2q+1)) w_x^{2q+1} w_t
+\tfrac{1}{2}b w_x^{q}w_t w_y
, 
\end{aligned}&&
\\
\quad&\begin{aligned}
Y_{(3)}= & 
\sigma_2 w_t w_y 
+\tfrac{1}{2} (b/(q+1)) w_x^{q+1} w_t
; 
\end{aligned}&&
\end{flalign}
\end{subequations}
\begin{subequations}\label{conslaw4}
\begin{flalign}
\quad&\begin{aligned}
T_{(4)} = &
\tfrac{1}{2} w_x w_y
, 
\end{aligned}&&
\\
\quad&\begin{aligned}
X_{(4)} = & 
w_y w_{xxx}
-w_{xy}w_{xx}
+\tfrac{b}{2} w_x^q w_y^2
+(\sigma_1/(2q+1))w_x^{2q+1} w_y
+\tfrac{1}{2} w_t w_y
, 
\end{aligned}&&
\\
\quad&\begin{aligned}
Y_{(4)}= & 
\tfrac{1}{2} w_{xx}^2
+\tfrac{1}{2}\sigma_2 w_y^2
-\tfrac{1}{2}(\sigma_1/((q+1)(2q+1))) w_x^{2(q+1)}
-\tfrac{1}{2} w_t w_x
; 
\end{aligned}&&
\end{flalign}
\end{subequations}

\item $q=\tfrac{1}{2}$, $a=\tfrac{1}{4}b$
\begin{subequations}\label{conslaw5}
\begin{flalign}
\quad&\begin{aligned}
T_{(5)} = &
\big(
\tfrac{3}{2}w_{xx}^2
-\tfrac{1}{2}\sigma_1 w_x^3
-b w_x^{3/2} w_y
-\tfrac{3}{2}\sigma_2 w_y^2
\big)t
+\tfrac{1}{2}x w_x^2
+y w_x w_y
, 
\end{aligned}&&
\\
\quad&\begin{aligned}
X_{(5)} = & 
\big( 
3 w_t w_{xxx}
-3 w_{tx}w_{xx}
+\tfrac{3}{2}w_t^2
+\tfrac{3}{2}b w_x^{1/2} w_t w_y
+\tfrac{3}{2}\sigma_1 w_x^2 w_t 
\big)t 
\\&\quad
+\big( 
w_x w_{xxx}
-\tfrac{1}{2} w_{xx}^2
+\tfrac{1}{3}\sigma_1 w_x^3
+\tfrac{1}{6}b w_x^{3/2} w_y
-\tfrac{1}{2}\sigma_2 w_y^2
\big)x
\\&\quad
+\big( 
b w_x^{1/2} w_y^2
+\sigma_1 w_x^2 w_y
+ w_t w_y
+2 w_y w_{xxx} 
-2 w_{xy}w_{xx}
\big)y
\\&\quad
+w w_{xxx} 
-2w_x w_{xx}
+\tfrac{1}{2}\sigma_1 w w_x^2
+\tfrac{1}{2} b w_x^{1/2} w_y w
+w w_t
, 
\end{aligned}&&
\\
\quad&\begin{aligned}
Y_{(5)}= & 
\big(
b w_t w_x^{3/2}
+3\sigma_2 w_t w_y
\big)t
+ 
\big(
\tfrac{1}{3}b w_x^{5/2}
+\sigma_2 w_x w_y
\big)x
\\&\quad
+\big(
w_{xx}^2
-\tfrac{1}{3}\sigma_1 w_x^3
+\sigma_2 w_y^2
-w_t w_x
\big)y
+\tfrac{1}{3}b w w_x^{3/2}
+\sigma_2 w w_y
; 
\end{aligned}&&
\end{flalign}
\end{subequations}

\item $q=\tfrac{1}{2}$, $a=0$
\begin{subequations}\label{conslaw6}
\begin{flalign}
\quad&\begin{aligned}
T_{(6)} = &
-\tfrac{1}{2}\sigma_1 t w_x^2 +x w_x
, 
\end{aligned}&&
\\
\quad&\begin{aligned}
X_{(6)} = & 
\big(
\tfrac{1}{2}\sigma_1 w_{xx}^2
-\tfrac{1}{3} w_x^3
+\tfrac{1}{2}\sigma_1\sigma_2 w_y^2
-\sigma_1 w_x w_{xxx}
\big)t
+\big(
\tfrac{1}{2}\sigma_1 w_x^2
+w_{xxx}
\big)x
- w_{xx}
, 
\end{aligned}&&
\\
\quad&\begin{aligned}
Y_{(6)}= & 
-\big(
\sigma_1\sigma_2 w_x w_y
+\tfrac{2}{5}\sigma_1 bw_x^{5/2}
\big)t
+\big(
\sigma_2 w_y
+\tfrac{2}{3}b w_x^{3/2}
\big)x
;
\end{aligned}&&
\end{flalign}
\end{subequations}

\item $q=1$, $a=\tfrac{1}{2}b$
\begin{subequations}\label{conslaw7}
\begin{flalign}
\quad&\begin{aligned}
T_{(7)} = &
(\tfrac{4}{3}\sigma_1\sigma_2 -\tfrac{1}{2}b^2) t w_x w_y
+bx w_x
-\tfrac{2}{3}\sigma_1 y w_x^2
, 
\end{aligned}&&
\\
\quad&\begin{aligned}
X_{(7)} = & 
(\tfrac{8}{3}\sigma_1\sigma_2 -b^2) 
\big(
w_y w_{xxx} - w_{xy} w_{xx}
+\tfrac{1}{2}w_t w_y
+\tfrac{1}{3}\sigma_1 w_x^3 w_y
+\tfrac{1}{2}b w_x w_y^2
\big) t
\\&\quad
+\big(
\tfrac{2}{3}\sigma_1 w_{xx}^2
+\tfrac{2}{3}\sigma_1\sigma_2 w_y^2
-\tfrac{1}{3} w_x^4
-\tfrac{1}{3}\sigma_1 bw_x^2 w_y
-\tfrac{4}{3}\sigma_1 w_x w_{xxx}
\big)y
\\&\quad
+b\big(
w_{xxx}
+\tfrac{1}{3}\sigma_1 w_x^3
+\tfrac{1}{2}b w_x w_y
\big)x
-b w_{xx}
, 
\end{aligned}&&
\\
\quad&\begin{aligned}
Y_{(7)}= & 
(\tfrac{8}{3}\sigma_1\sigma_2 -b^2) 
\big(
\tfrac{1}{2}w_{xx}^2
-\tfrac{1}{2} w_t w_x
-\tfrac{1}{12}\sigma_1 w_x^4
+\tfrac{1}{2}\sigma_2 w_y^2
\big)t
+b\big(
\sigma_2 w_y
+\tfrac{1}{4}b w_x^2
\big)x
\\&\quad
-\big(
\tfrac{4}{3}\sigma_1\sigma_2 w_x w_y
+\tfrac{1}{3}b\sigma_1 w_x^3
\big)y  
;
\end{aligned}&&
\end{flalign}
\end{subequations}

\item $q=1$, $b=\tfrac{1}{2}a+ \sigma_1\sigma_2/a$
\begin{subequations}\label{conslaw8}
\begin{flalign}
\quad&\begin{aligned}
T_{(8)} = &
-\tfrac{1}{2}a\sigma_2 y w_x^2 +x w_x
, 
\end{aligned}&&
\\
\quad&\begin{aligned}
X_{(8)} = & 
\big(
\tfrac{1}{2}\sigma_2 a w_{xx}^2
-\tfrac{1}{4}\sigma_1\sigma_2 a w_x^4
-\tfrac{1}{2}\sigma_2 a^2 w_x^2 w_y
+\tfrac{1}{2}a w_y^2
-\sigma_2 a w_x w_{xxx}
\big)y
\\&\quad
+\big(
\tfrac{1}{3}\sigma_1 w_x^3
+a w_x w_y
+w_{xxx}
\big)x
-w_{xx}
, 
\end{aligned}&&
\\
\quad&\begin{aligned}
Y_{(8)}= & 
\big(
\sigma_2 w_y
-\tfrac{1}{4}(a-2\sigma_1\sigma_2/a) w_x^2
\big)x
-\big(
\tfrac{1}{3}\sigma_1 w_x^3 +a w_x w_y
\big)y
;
\end{aligned}&&
\end{flalign}
\end{subequations}

\item $q=1$, $b^2=a^2+2\sigma_1\sigma_2$
\begin{subequations}\label{conslaw9}
\begin{flalign}
\quad&\begin{aligned}
T_{(9)} = &
\tfrac{3}{2}(b-a)w_{xx}^2
+\tfrac{1}{4}\sigma_1(a-b)w_x^4
-\tfrac{1}{2}\sigma_2(a+b)w_y^2 
-\sigma_1\sigma_2 w_x^2 w_y
,
\end{aligned}&&
\\
\quad&\begin{aligned}
X_{(9)} = & 
(2a-b)\big(
w_{xxx}^2
+\tfrac{2}{3}\sigma_1 w_x^3 w_{xxx}
+(a+b) w_x w_y w_{xxx}
+2\sigma_2 w_{xx} w_{yy}
+\sigma_2 w_{xy}^2
\\&\quad
+(a-b)( \tfrac{1}{2} w_y w_{xx}^2 -2 w_x w_{xy} w_{xx} )
+\tfrac{1}{9} w_x^6
+\tfrac{1}{2}a(a+b)w_x^2w_y^2
\big)
\\&\quad
+(a+b)\big(
w_t w_{xxx}
+\tfrac{1}{2} w_t^2
+\tfrac{1}{3}\sigma_1 w_t w_x^3
+a w_t w_x w_y
\big)
+3(a-b) w_{tx} w_{xx}
\\&\quad
+\tfrac{1}{12}(a(7a+b)\sigma_1-6\sigma_2) w_x^4 w_y
+\tfrac{1}{6}(2\sigma_1-a(a+b)\sigma_2) w_y^3
, 
\end{aligned}&&
\\
\quad&\begin{aligned}
Y_{(9)}= & 
(2a-b)\big(
\tfrac{1}{2}(a-b) w_x w_{xx}^2
-2\sigma_2 w_{xy} w_{xx}
-\tfrac{1}{3}(a^2-b^2) w_x^3 w_y
\big)
\\&\quad
-\tfrac{1}{2}(a^2-b^2) w_t w_x^2
+\sigma_2(a+b) w_t w_y
+\tfrac{1}{12}(3a(b-a)\sigma_1 -2\sigma_2) w_x^5
\\&\quad
+\tfrac{1}{2}(a(a+b)\sigma_2-2\sigma_1) w_x w_y^2
;  
\end{aligned}&&
\end{flalign}
\end{subequations}

\item $q=-2$, $a=-b$
\begin{subequations}\label{conslaw10}
\begin{flalign}
\quad&\begin{aligned}
T_{(10)} = &
\tfrac{1}{2} y w_x^2 -\sigma_2 t w_x w_y
, 
\end{aligned}&&
\\
\quad&\begin{aligned}
X_{(10)} = & 
\big(
\tfrac{2}{3}\sigma_1\sigma_2 w_y w_x^{-3}
-\sigma_2(2w_y w_{xxx} -2 w_{xy}w_{xx} + w_t w_y +b w_y w_x^{-2})
\big)t 
\\&\quad
+\big( 
w_x w_{xxx}
-\tfrac{1}{2}w_{xx}^2
-\tfrac{1}{2}\sigma_2 w_y^2
+b w_y w_x^{-1}
-\tfrac{1}{2}\sigma_1 w_x^{-2}
\big)y
, 
\end{aligned}&&
\\
\quad&\begin{aligned}
Y_{(10)}= & 
\big(
\sigma_2(w_t w_x - w_{xx}^2) 
- w_y^2
+\tfrac{1}{3}\sigma_1\sigma_2 w_x^{-2}
\big)t
+\sigma_2 y w_x w_y
; 
\end{aligned}&&
\end{flalign}
\end{subequations}
\enumnoindent
and, with an arbitrary function $f(t)$,\\

\item $a=bq$
\begin{subequations}\label{conslaw11}
\begin{flalign}
\quad&\begin{aligned}
T_{(11)} = &
yw_x f(t)
, 
\end{aligned}&&
\\
\quad&\begin{aligned}
X_{(11)} = & 
\big(
w_{xxx}
+b w_x^q w_y 
+(\sigma_1/(2q+1))w_x^{2q+1} 
\big) y f(t)
-w y f'(t)
, 
\end{aligned}&&
\\
\quad&\begin{aligned}
Y_{(11)}= & 
\sigma_2 \big(y w_y -w\big)  f(t)
; 
\end{aligned}&&
\end{flalign}
\end{subequations}

\item $q=1$
\begin{subequations}\label{conslaw12}
\begin{flalign}
\quad&\begin{aligned}
T_{(12)} = &
\tfrac{1}{2} (a-b) w_x^2 f(t)
, 
\end{aligned}&&
\\
\quad&\begin{aligned}
X_{(12)} = & 
(a-b)\big(
w_x w_{xxx}
-\tfrac{1}{2} w_{xx}^2
+\tfrac{1}{4}\sigma_1 w_x^4
+\tfrac{1}{2}a w_x^2 w_y
-\tfrac{1}{2}\sigma_2 w_y^2
\big) f(t)
\\&\quad
+\big(
w_{xxx}
+\tfrac{1}{3}\sigma_1 w_x^3
+a w_x w_y
+w_t
\big) y f'(t)
, 
\end{aligned}&&
\\
\quad&\begin{aligned}
Y_{(12)}= & 
\big(
(\tfrac{1}{6}(a-b)(2b-a) w_x^3
+\sigma_2 (a-b) w_x w_y
\big) f(t) 
+\big(
\tfrac{1}{2}(b-a) w_x^2
+\sigma_2 w_y)y
-\sigma_2 w \big) f'(t)
;
\end{aligned}&&
\end{flalign}
\end{subequations}

\item $q=1$, $a=\tfrac{1}{2}b$, $b^2=-\tfrac{8}{3}\sigma_1\sigma_2$
\begin{subequations}\label{conslaw13}
\begin{flalign}
\quad&\begin{aligned}
T_{(13)}= &
\tfrac{1}{2}w_{y}w_{x}f(t)
+\sigma_2(\tfrac{3}{16}\sigma_1 bw_{x} x -\tfrac{1}{8} w_{x}^2 y)f'(t)
,
\end{aligned}&&
\\
\quad&\begin{aligned}
X_{(13)} = & 
\big(
w_{y}w_{xxx}-w_{xy}w_{xx}
+\tfrac{1}{2} b w_{y}^2w_{x}
+\tfrac{1}{3}\sigma_1 w_{y}w_{x}^3
+\tfrac{1}{2} w_{t}w_{y}
\big)f(t)
\\&\quad
+\big(
{-\tfrac{3}{16}}\sigma_1 \sigma_2 bw_{xx}
+(\tfrac{3}{16}\sigma_1 \sigma_2 bw_{xxx}
+\tfrac{1}{16}\sigma_2 b w_{x}^3
-\tfrac{1}{4} w_{x}w_{y} )x
\\&\quad
+(\tfrac{1}{8}\sigma_2 w_{xx}^2
-\tfrac{1}{16}\sigma_2 \sigma_1 w_{x}^4
-\tfrac{1}{16}\sigma_2 b w_{x}^2 w_{y}
+\tfrac{1}{8}w_{y}^2
-\tfrac{1}{4}\sigma_2 w_{x}w_{xxx})y
\big)f'(t)
\\&\quad
+\big(
{-\tfrac{3}{32}}\sigma_1 b( w_{t} +w_{xxx} )
-\tfrac{1}{32}b w_{x}^3
+\tfrac{1}{8}\sigma_2 w_{y}w_{x}
\big)y^2 f''(t)
,
\end{aligned}&&
\\
\quad&\begin{aligned}
Y_{(13)} = & 
\big(
\tfrac{1}{2}w_{xx}^2
+\tfrac{1}{2}\sigma_2 w_{y}^2
-\tfrac{1}{12}\sigma_1 w_{x}^4
-\tfrac{1}{2}w_{t}w_{x}
\big)f(t)
\\&\quad
+\big(
(\tfrac{3}{16}\sigma_1 bw_{y}-\tfrac{1}{8}w_{x}^2)x
+(-\tfrac{1}{4}w_{x}w_{y}-\tfrac{1}{16}\sigma_2 bw_{x}^3)y
\big)f'(t)
\\&\quad
+\big(
(\tfrac{3}{16}\sigma_1 \sigma_2 b w) y
-(\tfrac{3}{32}\sigma_1 \sigma_2 b w_{y} -\tfrac{1}{16}\sigma_2 w_{x}^2)y^2
\big)f''(t)
;
\end{aligned}&&
\end{flalign}
\end{subequations}

\item $q=1$, $b=0$, $a^2=-2\sigma_1\sigma_2$
\begin{subequations}\label{conslaw14}
\begin{flalign}
\quad&\begin{aligned}
T_{(14)}= &
\big(
\tfrac{1}{2}w_{x}^2 y +\tfrac{1}{2}\sigma_1 a w_x x
\big)f(t)
,
\end{aligned}&&
\\
\quad&\begin{aligned}
X_{(14)} = & 
\big(
{-\tfrac{1}{2}}a\sigma_1 w_{xx}
+(\tfrac{1}{6}aw_{x}^3
-\sigma_2 w_{x}w_{y}
+\tfrac{1}{2}\sigma_1 aw_{xxx} )x
\\&\quad
+(w_{x}w_{xxx}
-\tfrac{1}{2}w_{xx}^2
+\tfrac{1}{4}\sigma_1 w_{x}^4
+\tfrac{1}{2}aw_{x}^2w_{y}
-\tfrac{1}{2}\sigma_2 w_{y}^2)y
\big)f(t)
\\&\quad
-\big(
\tfrac{1}{4}\sigma_2 \sigma_1 a w_{t} 
+\tfrac{1}{12}\sigma_2 a w_{x}^3
-\tfrac{1}{2}w_{x}w_{y}
+\tfrac{1}{4}\sigma_2 \sigma_1 a w_{xxx}
\big) y^2 f'(t)
,
\end{aligned}&&
\\
\quad&\begin{aligned}
Y_{(14)} = & 
\big(
\tfrac{1}{2}\sigma_2 (\sigma_1 a w_{y}+ w_{x}^2)x
+(\sigma_2 w_{x} w_{y} -\tfrac{1}{6}aw_{x}^3)y
\big)f(t)
\\&\quad
+\big(
\tfrac{1}{2}\sigma_1 a w y
-\tfrac{1}{4}(\sigma_1 a w_{y} +w_{x}^2))y^2
\big)f'(t)
;
\end{aligned}&&
\end{flalign}
\end{subequations}

\begin{subequations}\label{conslaw15}
\begin{flalign}
\quad&\begin{aligned}
T_{(15)}= &
\big(
{-\tfrac{9}{8}}\sigma_1 w_{xx}^2
+\tfrac{3}{16}w_{x}^4
+\tfrac{3}{8}\sigma_1(aw_{x}^2w_{y} -\sigma_2 w_{y}^2)
\big)f(t)
-\tfrac{3}{8} w_{x}^2 x f'(t)
\\&\quad
+\big(
\tfrac{3}{16}\sigma_1\sigma_2 w_{x}^2 y^2 -\tfrac{3}{8}\sigma_2 aw y
\big)f''(t)
,
\end{aligned}&&
\\
\quad&\begin{aligned}
X_{(15)} = & 
\big(
\tfrac{3}{2}\sigma_1 w_{xxx}^2
+(\tfrac{3}{4}\sigma_1 w_{t}+w_{x}^3+\tfrac{3}{2}a\sigma_1 w_{y}w_{x})w_{xxx}
+\tfrac{3}{4}\sigma_1 aw_{xx}^2w_{y}
+\tfrac{3}{2}\sigma_1 \sigma_2 w_{xy}^2
+\tfrac{3}{8}\sigma_1 w_{t}^2
\\&\quad
+(3\sigma_1\sigma_2 w_{yy}-\tfrac{3}{2}a\sigma_1 w_{x}w_{xy}
+\tfrac{9}{4}\sigma_1 w_{tx})w_{xx}
+(\tfrac{3}{4}\sigma_1 a w_{y}w_{x}+\tfrac{1}{4}w_{x}^3)w_{t}
\\&\quad
-\tfrac{1}{4}\sigma_1 \sigma_2 a w_{y}^3
+\tfrac{1}{6}\sigma_1 w_{x}^6
+\tfrac{5}{8}a w_{x}^4w_{y}
-\tfrac{3}{2}\sigma_2 w_{y}^2w_{x}^2
\big)f(t)
\\&\quad
+\big(
\tfrac{3}{4}\sigma_1(-w_{x}w_{xxx} +2 w_{xx}^2 + w_{x}w_{xx})
+(\tfrac{3}{8}\sigma_1 \sigma_2 w_{y}^2
-\tfrac{3}{8}\sigma_1 aw_{x}^2w_{y}
-\tfrac{3}{16}w_{x}^4)x
\big)f'(t)
\\&\quad
+\big(
{-\tfrac{3}{8}}\sigma_2 a w_{xx} y
+(\tfrac{1}{8}\sigma_1 \sigma_2 aw_{x}^3
+\tfrac{3}{8}\sigma_2 a(w_{xxx} +w_{t})
-\tfrac{3}{4}\sigma_1 w_{x}w_{y})yx
\\&\quad
+(\tfrac{3}{16}\sigma_1 \sigma_2(2 w_{x}w_{xxx} -w_{xx}^2 +aw_{x}^2w_{y})
-\tfrac{3}{16}\sigma_1 w_{y}^2
+\tfrac{3}{32}\sigma_2 w_{x}^4)y^2
\big)f''(t)
\\&\quad
+\big(
-\tfrac{1}{48}\sigma_1 a w_{x}^3
+\tfrac{1}{8}\sigma_1 \sigma_2 w_{x}w_{y}
-\tfrac{1}{16}a(w_{xxx} +w_{t})
\big)y^3 f'''(t)
,
\end{aligned}&&
\\
\quad&\begin{aligned}
Y_{(15)} = & 
\big(
\tfrac{3}{4}\sigma_1 a w_{x}w_{xx}^2
-3\sigma_1 \sigma_2 w_{xx}w_{xy}
+\tfrac{3}{4}\sigma_1 \sigma_2 aw_{x} w_{y}^2
+\sigma_2 w_{x}^3w_{y}
-\tfrac{1}{8}a w_{x}^5
\\&\quad
+(\tfrac{3}{4}\sigma_1 \sigma_2 w_{y} -\tfrac{3}{8}\sigma_1a w_{x}^2)w_{t}
\big)f(t)
+\big(
-\tfrac{3}{4}\sigma_1 \sigma_2 w_{x}w_{y}
+\tfrac{1}{8}\sigma_1 a w_{x}^3
\big)x f'(t)
\\&\quad
+\big(
{-\tfrac{3}{8}}a w x
(\tfrac{3}{8}aw_{y} +\tfrac{3}{8}\sigma_1 w_{x}^2)xy
+(\tfrac{3}{8}\sigma_1 w_{x}w_{y}
-\tfrac{1}{16}\sigma_1 \sigma_2 a w_{x}^3)y^2
\big)f''(t)
\\&\quad
+\big(
\tfrac{3}{16}\sigma_2 a w y^2 
-(\tfrac{1}{16}\sigma_2 a w_{y} +\tfrac{1}{16}\sigma_1 \sigma_2 w_{x}^2)y^3
\big)f'''(t)
.
\end{aligned}&&
\end{flalign}
\end{subequations}
\end{enumerate}
\end{thm}

The physical meaning and properties of these conservation law will be discussed next.
Each conservation law corresponds to a conserved integral \eqref{conservedintegral}
holding in any given spatial domain $\Omega\subseteq\Rnum^2$
for all solutions $u(t,x,y)$ of the modified gKP equation \eqref{mgKP-scaled}.
Recall that the potential $w$ can be formally expressed in terms of $u$
via the relation \eqref{w-u-rel}.

\subsection{Conserved integrals in the non-variational case}

We will first consider the conservation laws \eqref{conslaw1}, \eqref{conslaw6},
\eqref{conslaw8}, \eqref{conslaw9}, 
which hold without the need for any variational structure,
namely when $a\neq \tfrac{1}{2}bq$, 
and which do not contain $f(t)$.

The first conservation law \eqref{conslaw1} yields the conserved integral 
\begin{equation}\label{mgKP-L2}
\mathcal{P}[u] = 
\tfrac{1}{2} \int_{\Omega} u^2\,dx\,dy .
\end{equation}
This is a momentum quantity, 
in analogy with the same conserved integral for the mKdV equation.
It shows that the $L^2$-norm of solutions $u(t,x,y)$ is conserved.

From conservation law \eqref{conslaw6}, we obtain a similar conserved integral
\begin{equation}\label{mgKP-Gal-L2}
\mathcal{G}[u] = 
\tfrac{1}{2}\int_{\Omega} \big( xu -\sigma_1 t u^2 \big)\,dx\,dy ,
\quad
q=\tfrac{1}{2},
\quad
a=0 .
\end{equation}
This is a Galilean momentum quantity.
It is related to the momentum \eqref{mgKP-L2} by
$\mathcal{G} = \mathcal{X} -\sigma_1 t \mathcal{P}$
where
\begin{equation}\label{mgKP-x-L1}
\mathcal{X}[u] = 
\tfrac{1}{2}\int_{\Omega} x u \,dx\,dy 
\end{equation}
is an $x$-moment of mass. 
Up to boundary terms, we have 
\begin{equation}\label{x-L1-rel}
\frac{d}{dt}\mathcal{X}[u] = \sigma_1 \mathcal{P}[u] , 
\end{equation}
showing that the $x$-momentum of mass undergoes free particle motion.
A similar relation is well known to hold for the KdV equation.

The other two conservation laws \eqref{conslaw8} and \eqref{conslaw9}
yield the respective conserved integrals
\begin{equation}\label{mgKP-x-L1-y-L2}
\mathcal{C}[u] = 
\tfrac{1}{2}\int_{\Omega} (x u - \sigma_2 a y u^2 )\,dx\,dy ,
\quad
q=1,
\quad
b=\tfrac{1}{2}a+ \sigma_1\sigma_2/a , 
\end{equation}
and
\begin{equation}\label{mgKP-ener}
\mathcal{E}[u] =
(b-a)\int_{\Omega} \big(
\tfrac{3}{2}u_{x}^2 
-\tfrac{1}{4}\sigma_1 (u^2-(a+b)\partial_x^{-1}u_y)^2
\big)\,dx\,dy,
\quad
q=1,
\quad
b^2=a^2+2\sigma_1\sigma_2 . 
\end{equation}
This quantity \eqref{mgKP-ener} is an energy,
while the first quantity \eqref{mgKP-x-L1-y-L2}
is a linear combination of the $x$-moment of mass
and the $y$-moment of momentum
\begin{equation}\label{mgKP-y-L2}
\mathcal{Y}[u^2] = 
\tfrac{1}{2}\int_{\Omega} y u^2 \,dx\,dy . 
\end{equation}
As a consequence, we have
\begin{equation}\label{y-L2-rel}
\frac{d}{dt}\mathcal{Y}[u^2] = \sigma_1\sigma_2 \mathcal{P}/a , 
\end{equation}
whereby the $y$-moment of momentum undergoes the same motion
(up to a constant factor)
as the $x$-moment of mass. 

The mass itself appears to be missing here,
but it turns out to be a special case of a more general conservation law involving $f(t)$,
which we will discuss later.

\subsection{Conserved integrals in the variational case}

We will next consider the conservation laws
\eqref{conslaw3}--\eqref{conslaw5}, \eqref{conslaw7}, \eqref{conslaw10},
which hold only when the modified gKP equation \eqref{mgKP-scaled}
has a variational structure, namely $a =\tfrac{1}{2}bq$,
and which do not contain $f(t)$.
The associated multipliers
\eqref{multr3}--\eqref{multr5}, \eqref{multr7}, \eqref{multr10}
correspond to variational symmetries leaving the Lagrangian \eqref{mgKP-L}
invariant modulo a total divergence.

Conservation laws \eqref{conslaw3} and \eqref{conslaw4}
yield conserved integrals for energy and $y$-momentum 
\begin{gather}
\mathcal{E}_\var[u] = \tfrac{1}{2}\int_{\Omega}\big(
u_{x}^2 -\tfrac{1}{(q+1)(2q+1)} \sigma_1 (u^{q+1} +\tfrac{b(2q+1)}{2} \partial_x^{-1}u_y)^2 
-(\sigma_2 +\tfrac{b^2(2q+1)^2}{4} )(\partial_x^{-1}u_{y})^2 
\big)\,dx\,dy , 
\label{mgKP-var-ener}
\\
\mathcal{P}^y_\var[u] = \tfrac{1}{2} \int_{\Omega} 
u \partial_x^{-1}u_y
\,dx\,dy .
\label{mgKP-var-ymom}
\end{gather}
The associated multipliers \eqref{multr3} and \eqref{multr4}
correspond to a time-translation symmetry and a $y$-translation symmetry. 

From conservation law \eqref{conslaw5}, we obtain the conserved integral 
\begin{equation}
\mathcal{G}_\var[u]
= \int_{\Omega}\big(
t(
\tfrac{3}{2}u_{x}^2
-\tfrac{1}{2}\sigma_1 u^3
-b u^{3/2} \partial_x^{-1}u_y
-\tfrac{3}{2}\sigma_2 (\partial_x^{-1}u_y)^2
)
+\tfrac{1}{2}x u^2
+y u \partial_x^{-1}u_y
\big)\,dx\,dy ,
\quad
q=\tfrac{1}{2}, 
\end{equation}
which is a Galilean energy quantity.
It can be expressed as 
$\mathcal{G}_\var[u] = 3t \mathcal{E}_\var[u] +\mathcal{X}[u^2] + 2\mathcal{Y}[u \partial_x^{-1}u_y]$
in terms of the $x$-moment of momentum $\mathcal{X}[u^2]$ 
and the $y$-moment of $y$-momentum $\mathcal{Y}[u \partial_x^{-1}u_y]$. 
In particular, up to boundary terms, we have 
\begin{equation}\label{x-L2-rel}
\frac{d}{dt}\Big( \mathcal{X}[u^2] + 2\mathcal{Y}[u \partial_x^{-1}u_y] \big)
= -3\mathcal{E}_\var[u] .
\end{equation}
A similar relation is well known to hold for the mKdV equation.
The associated multiplier \eqref{multr5} corresponds to a scaling symmetry. 

The two other conservation laws \eqref{conslaw7} and \eqref{conslaw10} 
yield analogous conserved integrals
\begin{equation}
\begin{aligned}
\mathcal{C}_\var[u] & = \int_{\Omega}\big(
b x u
-\tfrac{2}{3}\sigma_1 y u^2
+(\tfrac{4}{3}\sigma_1\sigma_2 -\tfrac{1}{2}b^2) t u \partial_x^{-1}u_y
\big)\,dx\,dy ,
\quad
q=1
\\
& =2b\mathcal{X}[u] -\tfrac{4}{3}\sigma_1 \mathcal{Y}[u^2]
- (\tfrac{8}{3}\sigma_1\sigma_2 -b^2) t \mathcal{P}^y_\var[u] , 
\end{aligned}
\end{equation}
and
\begin{equation}
\begin{aligned}
\mathcal{C}_\var[u] & = \int_{\Omega}\big(
\tfrac{1}{2} y u^2 -\sigma_2 t u \partial_x^{-1}u_y
\big)\,dx\,dy ,
\quad
q=-2
\\
& = \mathcal{Y}[u^2] -2\sigma_2 t\mathcal{P}^y_\var . 
\end{aligned}
\end{equation}
which can be expressed in terms of the $x$-moment of mass \eqref{mgKP-x-L1}
and the $y$-moment of momentum.
Up to boundary terms, they respectively yield the similar relations 
\begin{equation}
\frac{d}{dt}\mathcal{Y}[u^2] =
\tfrac{3}{2}b \mathcal{P}[u] - (2\sigma_2 -\tfrac{3}{4}\sigma_1 b^2) \mathcal{P}^y_\var[u]
\end{equation}
after use of the Galilean momentum relation \eqref{x-L1-rel},
and
\begin{equation}
\frac{d}{dt}\mathcal{Y}[u^2] =
2\sigma_2 \mathcal{P}^y_\var[u] . 
\end{equation}
Their associated multipliers \eqref{multr7} and \eqref{multr10} 
respectively correspond to a boost in the $(x,y)$-plane
along a parabola $ y^2 +4\sigma_2 t x=\const$,
and a similar boost along 
$y^2-(\tfrac{3}{2}b^2-4\sigma_1\sigma_2)t x =\const$
combined with a boost to a moving frame with speed $b$,
namely, $by-(b^2-\tfrac{8}{3}\sigma_1\sigma_2)tu  =\const$,
with $u$ viewed as the wave speed.

\subsection{Conserved topological charges and integral constraints for the Cauchy problem} 

The conservation laws \eqref{conslaw2} and \eqref{conslaw11}--\eqref{conslaw15}
involve an arbitrary function $f(t)$.
As shown by the general results in \Ref{AncRec2020},
each of these conservation laws turns out to be
locally equivalent to a spatial flux conservation law
$f(t)(D_x X +D_y Y) =0$
holding for all solutions $u(x,y,t)$ of the modified gKP equation \eqref{mgKP-scaled}.
Note that $f(t)$ can then be dropped without loss of generality. 
The resulting conserved integral thereby has the form
\begin{equation}\label{charge}
\mathcal{Q}[u] = 
\oint_{\partial\Omega} {-Y}\,dx + X\,dy =0
\end{equation}
where $\partial\Omega$ is the closed boundary curve of a given spatial domain $\Omega\subseteq\Rnum^2$. 
This line integral holds without any boundary conditions on $u$
and it is unchanged under continuous deformations of the boundary curve. 
Therefore, it describes a conserved (time-independent) topological charge.

Any conservation law that involves $f(t)$
can be specialized by taking $f(t)$ to be a specific function of $t$.
This will yield a specific conservation law that is locally equivalent to a spatial flux conservation law.
Hereafter we will choose the constant function $f(t)=1$ . 

Conservation law \eqref{conslaw2} with $f(t)=1$
yields the conserved integral 
\begin{equation}
\mathcal{M}[u] = \int_{\Omega} u \,dx\,dy
= \oint_{\partial\Omega} \partial_x^{-1}u \,dy , 
\end{equation}
which is the mass of $u$. 
The equivalent topological charge is given by 
\begin{equation}\label{mgKP-charge-mass}
\oint_{\partial\Omega}
\big(\tfrac{a-bq}{q(q+1)} u^{q+1}-\sigma_2 \partial_x^{-1}u_y\big)\,dx
+ \big(\tfrac{1}{2q+1} \sigma_1 u^{2q+1} + u_{xx} +\tfrac{a}{q}u^{q} \partial_x^{-1}u_y +\partial_x^{-1}u_t\big)\,dy
=0 .
\end{equation}

Similarly, conservation law \eqref{conslaw11} with $f(t)=1$
yields the $y$-moment of mass
\begin{equation}
\mathcal{Y}[u] = \int_{\Omega} yu \,dx\,dy
= \oint_{\partial\Omega} y\partial_x^{-1}u \,dy,
\quad
a=bq
\end{equation}
whose conservation is equivalent to the topological charge
\begin{equation}\label{mgKP-charge-y-L2}
\oint_{\partial\Omega}
\big(
\sigma_2 (\partial_x^{-1}u -y\partial_x^{-1}u_y)
\big)\,dx
+ \big(
\tfrac{1}{2q+1} \sigma_1 u^{2q+1} + u_{xx} +bu^{q} \partial_x^{-1}u_y +\partial_x^{-1}u_t
\big)\,dy
=0,
\quad
a=bq . 
\end{equation}

Conservation laws \eqref{conslaw12} and \eqref{conslaw13} are more interesting.
For $f(t)=1$, they yield
the momentum \eqref{mgKP-L2} and the $y$-momentum \eqref{mgKP-var-ymom},
respectively. 
They are each equivalent to a topological charge, 
\begin{equation}
\begin{aligned}  
\oint_{\partial\Omega} & 
\big(
(b-a)(\sigma_2  u\partial_x^{-1}u_y +\tfrac{1}{6}(2 b-a) u^3)
-\sigma_2 \partial_x^{-1}u_t
+y((b-a) u u_{t} +\sigma_2 \partial_x^{-1}u_{ty})
\big)\,dx
\\&
+\big(
(a-b)(\tfrac{1}{4} \sigma_1  u^4 -\tfrac{1}{2} u_{x}^2+u u_{xx}
+\tfrac{1}{2} a u^2 \partial_x^{-1}u_y -\tfrac{1}{2} \sigma_2  (\partial_x^{-1}u_y)^2 )
\\&\quad
-y((\sigma_1   u^2 +a \partial_x^{-1}u_y) u_{t} + a u \partial_x^{-1}u_{ty} +\partial_x^{-1}u_{tt} + u_{txx})
\big)\,dy =0 ,
\quad
q=1 , 
\end{aligned}
\end{equation}
and
\begin{equation}
\begin{aligned}
\oint_{\partial\Omega} & 
\big(
{-\tfrac{1}{2}} u_{x}^2 +\tfrac{1}{12} \sigma_1   u^4
-\tfrac{1}{2} \sigma_2 (\partial_x^{-1}u_y)^2
+\tfrac{1}{2} u\partial_x^{-1}u_t 
+x( \tfrac{1}{4} u u_{t}-\tfrac{3}{16} \sigma_1  b \partial_x^{-1}u_{ty} )
\\&\quad
+y( (\tfrac{3}{16} \sigma_2   b u^2 + \tfrac{1}{4} \partial_x^{-1}u_y) u_{t}
+\tfrac{3}{16} \sigma_1   \sigma_2   b \partial_x^{-1}u_{tt} +\tfrac{1}{4} u \partial_x^{-1}u_{ty} )
\\&\quad
+y^2( \tfrac{1}{8} \sigma_2 (u_{t}^2 +u u_{tt}) -\tfrac{3}{32} \sigma_1   \sigma_2   b \partial_x^{-1}u_{tty} )
\big)\,dx
\\&
+\big(
\tfrac{1}{2} u_{x}^2 -\tfrac{1}{12} \sigma_1 u^4
-\tfrac{1}{2} \partial_x^{-1}u_t u
+\tfrac{1}{2} \sigma_2 (\partial_x^{-1}u_y)^2
+x( \tfrac{1}{4} u u_{t}-\tfrac{3}{16} \sigma_1  b \partial_x^{-1}u_{ty} )
\\&\quad
+y( (\tfrac{3}{16} \sigma_2 b u^2+\tfrac{1}{4} \partial_x^{-1}u_y) u_{t}
+\tfrac{1}{4} u \partial_x^{-1}u_{ty}
+\tfrac{3}{16} \sigma_1\sigma_2 b \partial_x^{-1}u_{tt} )
\\&\quad
+y^2( \tfrac{1}{8} \sigma_2 (u_{t}^2 +u u_{tt})
-\tfrac{3}{32} \sigma_1\sigma_2 b \partial_x^{-1}u_{tty} ) 
\big)\,dy =0 ,
\quad
q=1,\
a=\tfrac{1}{2}b,\
b^2=-\tfrac{8}{3}\sigma_1\sigma_2 . 
\end{aligned}
\end{equation}
From the equivalence,
the following integral relations can be shown to hold for all solutions $u(t,x,y)$:
\begin{equation}\label{mgKP-mom-rel}
\begin{aligned}
(a-b)\mathcal{P}[u] & = \tfrac{1}{2} (a-b) \int_{\Omega} u^2 \,dx\,dy
\\&
= \oint_{\partial\Omega}
\big(
y(\tfrac{1}{2}(b-a) u^2 +\sigma_2 \partial_x^{-1}u_y) -\sigma_2 \partial_x^{-1}u
\big)\,dx
\\&\qquad\quad
- \big( 
y(u_{xx}+\tfrac{1}{3} \sigma_1 u^3 +a u\partial_x^{-1}u_y +\partial_x^{-1}u_t)
\big)\,dy,
\quad
q=1 , 
\end{aligned}
\end{equation}
and 
\begin{equation}\label{mgKP-ymom-rel}
\begin{aligned}
\mathcal{P}^y[u] & = \tfrac{1}{2} \int_{\Omega} u\partial_x^{-1}u_y \,dx\,dy
\\&
= \oint_{\partial\Omega}
\big(
{-}x(\tfrac{1}{8} u^2 +\tfrac{3}{16} \sigma_2  b \partial_x^{-1}u_y) 
-y(\tfrac{1}{16} \sigma_2  b u^3
+\tfrac{1}{4} u \partial_x^{-1}u_y
-\tfrac{3}{16} b \partial_x^{-1}u_t)
\\&\qquad\qquad
-y^2(\tfrac{1}{8}\sigma_2  u u_{t} 
+\tfrac{3}{32} b\partial_x^{-1}u_{ty}) 
\big)\,dx 
\\&\qquad\quad
-\big(
\tfrac{3}{16} b u_{x}
+x(\tfrac{1}{16} \sigma_2  b u^3
-\tfrac{3}{16}  b u_{xx} 
-\tfrac{1}{4} u \partial_x^{-1}u_y
-\tfrac{3}{16}  b \partial_x^{-1}u_t )
\\&\qquad\qquad
+y(\tfrac{1}{16} u^4 
+\tfrac{1}{8} \sigma_2  u_{x}^2
-\tfrac{1}{4} \sigma_2 u u_{xx}
-\tfrac{1}{16} \sigma_2  b u^2 \partial_x^{-1}u_y
+\tfrac{1}{8} (\partial_x^{-1}u_y)^2) 
\\&\qquad\qquad
+y^2( (\tfrac{3}{32} b u^2 -\tfrac{1}{8} \sigma_2 \partial_x^{-1}u_y ) u_{t}
-\tfrac{3}{32} \sigma_2  b u_{txx}
\\&\qquad\qquad
-\tfrac{3}{32} \sigma_2  b \partial_x^{-1}u_{tt}
-\tfrac{1}{8} \sigma_2  u \partial_x^{-1}u_{ty}) 
\big)\,dy,
\quad
q=1,\
a=\tfrac{1}{2}b,\
b^2=-\tfrac{8}{3}\sigma_1\sigma_2 . 
\end{aligned}
\end{equation}
Thus, both the momentum and the $y$-momentum are line-integral quantities
which involve the values of $\partial_x^{-1}u$, $\partial_x^{-1}u_t$, and their spatial derivatives
evaluated only at the domain boundary $\partial\Omega$. 
This is a surprising result. 

A similar result can be derived for the energy \eqref{mgKP-ener}
and for the moment-quantity \eqref{mgKP-x-L1-y-L2}. 
when they are specialized to the case $b=0$ 
in which the modified gKP equation \eqref{mgKP-scaled} reduces to 
a slightly generalized version of the mKP equation with $\sigma_1\sigma_2 = -1$. 
The two conservation laws \eqref{conslaw14} and \eqref{conslaw15},
which hold in this case, 
give rise to topological charges that are respectively equivalent to
the energy and the moment-quantity.
This equivalence can be shown to yield the integral relation 
\begin{equation}\label{mgKP-ener-rel}
\begin{aligned}
\mathcal{E}[u] 
& = \int_{\Omega} \big(
\tfrac{3}{2}u_{x}^2 
+\tfrac{1}{4}\sigma_2 (u^2 -a\partial_x^{-1}u_y)^2 
\big)\,dx\,dy
\\
& = \oint_{\partial\Omega} 
\big(
x( a({-}\tfrac{1}{6}u^3 + \tfrac{1}{2}\sigma_2 \partial_x^{-1}u_t) 
+ \sigma_2 u \partial_x^{-1}u_y )
+ xy( u u_t - \tfrac{1}{2}\sigma_2 a \partial_x^{-1}u_{ty} )
\\&\qquad\qquad
+y^2( -\tfrac{1}{4}\sigma_2 a u^2 u_t 
+\tfrac{1}{2} u_t \partial_x^{-1}u_{y} +\tfrac{1}{2} u \partial_x^{-1}u_{ty} 
+\tfrac{1}{4} a \partial_x^{-1}u_{tt} )
\\&\qquad\qquad
+ y^3( \tfrac{1}{6} \sigma_2 u_t^2  +\tfrac{1}{6}\sigma_2 u u_{tt} 
-\tfrac{1}{12} a \partial_x^{-1}u_{tty} )
\big)\,dx
\\&\qquad\quad
+\big( 
u u_x 
+ x( \tfrac{1}{4} \sigma_2 u^4 +\tfrac{1}{2} u_x^2 -u u_{xx} +\sigma_2 (\partial_x^{-1}u_y)^3 -\tfrac{1}{2} a u^2 \partial_x^{-1}u_y )
\\&\qquad\qquad
+y( \tfrac{1}{2} a u_{tx} )
+ xy( -\tfrac{1}{2} \sigma_2 a u^2 u_t +\tfrac{1}{4} u_{txx}
+ u_t \partial_x^{-1}u_y + u \partial_x^{-1}u_{ty} 
+\tfrac{1}{4} \partial_x^{-1}u_{tt} )
\\&\qquad\qquad
+ y^2 ( \tfrac{1}{2} u^3 u_t 
-\sigma_2(u_{xx} u_t - u_x u_{tx} + u u_{txx})
-\sigma_2 a(\tfrac{1}{2} u u_t  \partial_x^{-1}u_y +\tfrac{1}{4} u^2 \partial_x^{-1}u_{ty}) 
\\&\qquad\qquad
+\tfrac{1}{2} \partial_x^{-1}u_{y} \partial_x^{-1}u_{ty} )
+ y^3( 
\tfrac{1}{12} a(2\sigma_2 u u_t^2 - u^2 u_{tt} +\sigma_2 u_{ttxx} -\partial_x^{-1}u_{ttt})  
\\&\qquad\qquad
+\tfrac{1}{6} \sigma_2( u_{tt}  \partial_x^{-1}u_{y} +2 u_t  \partial_x^{-1}u_{ty} + u \partial_x^{-1}u_{tty}) )
\big)\,dy,
\quad
q=1,\
b=0,\
a^2=-2\sigma_1\sigma_2
\end{aligned}
\end{equation}
which holds for all solutions $u(t,x,y)$. 

These integral relations \eqref{mgKP-mom-rel}, \eqref{mgKP-ymom-rel}, \eqref{mgKP-ener-rel}
hold in the case $q=1$, 
which corresponds to the general scaled form of the universal modified KP-like equation \eqref{genmKP}: 
\begin{equation}\label{scaled-genmKP}
u_t + (\sigma_1 u^{2} +a \partial_x^{-1}u_y) u_x + b u u_y+ u_{xxx} +\sigma_2 \partial_x^{-1} u_{yy} =0 .
\end{equation}
Hereafter we take $\Omega=\Rnum^2$. 
By examining the asymptotic conditions on $u(t,x,y)$ for which 
the integrals on the both sides of each relation will vanish, 
and using the general argument shown in \Ref{AncRec2020}, 
we obtain the following conclusions about the initial-value problem on $\Rnum^2$. 

First,
the $y$-momentum relation \eqref{mgKP-ymom-rel} gives rise to an integral constraint 
$\mathcal{P}^y[u]=0$ on initial-value solutions with spatial decay 
$u(t,x,y)=O(x^{-3}y^{-3})$ as $|x|,|y|\to \infty$. 

Second, 
the momentum relation \eqref{mgKP-mom-rel} gives rise to an integral constraint 
$\mathcal{P}[u]=0$ on initial-value solutions with spatial decay 
$u(t,x,y)=O(x^{-2}y^{-2})$ as $|x|,|y|\to \infty$. 
When $b\neq a$, this constraint implies that $\| u\|_{L^2}=0$, 
whereby $u=0$ would be the only possible solution. 
Thus, in this case the initial-value problem is ill-posed in $L^2$. 
When $b=a$, the constraint requires that $u(t,x,y)=O(x^{-2}y^{-2})$ as $|x|,|y|\to \infty$, 
since otherwise the initial-value will be ill-posed in $L^2$. 

Third, 
the energy relation \eqref{mgKP-ener-rel} gives rise to an integral constraint 
$\mathcal{E}[u]=0$ on initial-value solutions with spatial decay 
$u(t,x,y)=O(x^{-3/2}y^{-4})$ as $|x|,|y|\to \infty$. 
In the case $\sigma_2=1$, the energy integral (on the left side) is non-negative,
and thus the constraint leads to $u=0$. 
As a result, the initial-value problem is ill-posed in the energy space. 
In the opposite case $\sigma_2 =-1$, 
since the energy integral has indefinite sign, 
well-posedness for solutions with the decay 
$u(t,x,y)=O(x^{-3/2}y^{-4})$ as $|x|,|y|\to \infty$
requires that the initial data has zero energy. 

\begin{prop}
For the universal modified KP-like equation \eqref{scaled-genmKP}:\\
(i) Well-posedness of solutions with spatial decay 
$u|_{t=0} =O(x^{-3}y^{-3})$ as $|x|,|y|\to \infty$ 
can hold only if the initial data satisfies the constraint $\int_{\Rnum^2} u\partial_x^{-1} u_y\,dx\,dy=0$. 
\\
(ii) Well-posedness can hold in $L^2$ only if $b=a$ and only if 
initial data has the spatial decay $u|_{t=0}=O(x^{-2}y^{-2})$ as $|x|,|y|\to \infty$. 
\\
(iii) Well-posedness can hold in the energy space only if $\sigma_1=-\sigma_2=1$ 
and only if the initial data has zero-energy  and spatial decay 
$u|_{t=0}=O(x^{-3/2}y^{-4})$ as $|x|,|y|\to \infty$. 
\end{prop}

\subsection{Critical powers and sign properties}

Since conserved energy quantities play a crucial role
in the global analysis of solutions,
we will examine the sign property of
the energy \eqref{mgKP-var-ener} in the variational case
and the energy \eqref{mgKP-ener} in the non-variational case.
From these expressions, we see that 
$\mathcal{E}[u]\geq 0$ if $\sigma_1=-1$
and 
$\mathcal{E}_\var[u]\geq 0$ if $\sigma_1=\sigma_2=-1$.
Hence the non-variational energy is non-negative in the defocussing case,
while the variational energy is non-negative in the defocussing case with sign-changing dispersion. 

Another important feature of the energy is its behaviour
under the scaling symmetry \eqref{mgKP-pot-scal}
of the modified gKP equation \eqref{mgKP-scaled}. 

In general,
when a conserved integral \eqref{conservedintegral} is scaling homogeneous,
it will have a scaling weight $d$ defined by 
$\mathcal{C}[u]\to \lambda^d \mathcal{C}[u]$
under the symmetry \eqref{mgKP-pot-scal}. 
If $d<0$ or $d>0$,
then the conserved integral is said to be \emph{subcritical} or \emph{supercritical}, respectively.
The \emph{critical} case $d=0$ corresponds to the conserved integral being scaling invariant.
Typically, global (long-time) existence of solutions to the Cauchy problem
can be established 
for all initial data without any condition on the size of the energy
in the subcritical case,
and at least for initial data with sufficiently small energy
in the critical case.

The scaling weight and criticality of the momentum ($L^2$ norm) and the energies
is shown in table~\ref{table:criticality}.
We see that the momentum is subcritical for $q<\tfrac{2}{3}$, 
while the energy is subcritical for $q<2$.
Therefore, when $q$ is an integer or a half-integer,
global existence can be expected to hold in $L^2$ only for $q=\tfrac{1}{2}$
and in the energy space only for $q=\tfrac{1}{2},1,\tfrac{3}{2},2$. 

\begin{table}[h!]
\caption{Scaling properties of momentum and energy}
\label{table:criticality}
\centering
\begin{tabular}{l|c||c|c}
\hline
Conserved integral
& $q$
& Scaling weight
& Criticality
\\
\hline
\hline
momentum $\mathcal{P}[u]$
&
$>0$
&
$3-2/q$
&
$q=\tfrac{2}{3}$
\\
\hline
energy $\mathcal{E}_\var[u]$
&
$>0$
&
$1-2/q$
&
$q=2$ 
\\
\hline
energy $\mathcal{E}[u]$
&
$1$
&
$-1$
&
subcritical
\\
\hline
\end{tabular}
\end{table}

\section{Line-soliton and line-shock solutions}\label{sec:solns}

A line travelling wave in two dimensions has the form 
\begin{equation}\label{linesoliton}
u=U(\xi), 
\quad
\xi =x+\mu y-\nu t
\end{equation}
where the parameters $\mu$ and $\nu$ determine the direction and the speed of the wave.
The amplitude of a line travelling wave is translation-invariant in the perpendicular direction. 
If the amplitude exhibits exponential asymptotic decay for large $|\xi|$, 
then the travelling wave is a \emph{line-soliton}. 

As noted previously in \Ref{AncGanRec2018,AncGanRec2019},  
a more geometrical form for a line travelling wave is given by writing 
$x+\mu y= (x,y)\cdot\kvec$ 
with $\kvec=(1,\mu)$ being a constant vector in the $(x,y)$-plane. 
The travelling wave variable can then be expressed as 
\begin{equation}\label{travwavevar}
\xi = |\kvec|( \hat\kvec\cdot (x,y) - c t )
\end{equation}
where the unit vector 
\begin{equation}
\hat\kvec = (\cos\theta,\sin\theta),
\quad
\tan\theta = \mu
\end{equation}
gives the direction of propagation of the wave, 
and the constant 
\begin{equation}
c = \nu/|\kvec|, 
\quad
|\kvec|^2 = 1+\mu^2
\end{equation}
gives the speed of the wave. 
We will take the domain of $\theta$ to be $-\tfrac{1}{2}\pi < \theta \leq \tfrac{1}{2}\pi$,
since the direction of propagation stays the same 
when both the direction angle is changed by $\pm\pi$ 
and the sign of the speed is reversed.

We will now derive the explicit line-soliton solutions \eqref{linesoliton}
for the modified gKP equation in the scaled form \eqref{mgKP-scaled}
when $p=2q$ is a positive integer. 
As main results, the nature of the solutions is shown to depend 
essentially on whether $q$ is an even integer, an odd integer, or a half-integer:
symmetrical bright/dark pairs of line-solitons are admitted
in the even case; 
non-symmetrical bright/dark pairs of line-solitons are admitted 
in the odd case when $\alpha/\beta >0$;
single line-solitons are admitted in the odd case when $\alpha/\beta <0$
and also in the half-integer case.
A line-shock solution is shown to arise under a certain condition relating $\mu$, $\nu$, $q$, and the coefficients in the equation.

\subsection{Derivation}

It will be convenient to use the coordinate expression for the travelling wave variable
$\xi = x+\mu y-\nu t$,
so thus $u_x = U'$, $u_y=\mu U'$, $u_t = -\nu U'$, and so on, 
while $\partial_x^{-1}u_y = \mu\partial_x^{-1} U'$
depends on the choice of asymptotic condition on $U$
through the constants $x_1$ and $x_2$ in the relation \eqref{w-u-rel}
which defines $\partial_x^{-1}$.

We will be interested in travelling waves whose amplitude $U$
vanishes as $\xi\to -\infty$.
Thus, we take $x_2=x_1=-\infty$, whereby 
\begin{equation}
\partial_x^{-1}u_y = \mu\partial_x^{-1} U' = 
\mu \int^\xi_{-\infty} U'(x)\,dx  = \mu U . 
\end{equation}

Substitution of the line-soliton expression \eqref{linesoliton} into equation \eqref{mgKP-scaled} 
yields a nonlinear third-order ODE
\begin{equation}\label{mgKP-U-linesoliton-ode}
(\sigma_2\mu^2 -\nu) U'+ (\sigma_1 U^{2q} +(a+b)\mu U^q) U' + U'''=0 .
\end{equation}
This ODE can be integrated to quadrature by starting from 
the conservation laws \eqref{conslaw1} for momentum and \eqref{conslaw2} for mass
of the equation \eqref{mgKP-scaled}. 
The method is based on symmetry multi-reduction \cite{AncGan2020}
utilizing the travelling wave symmetries generated by 
$\X_1 = (\mu^2+1)\partial_t + \nu\partial_x + \nu\mu \partial_y$
and $\X_2 = \mu\partial_x - \partial_y$.
These two symmetries form an abelian algebra whose invariants are $\xi$ and $u$. 
In particular, 
reduction of equation \eqref{mgKP-scaled} under this symmetry algebra
yields the ODE \eqref{mgKP-U-linesoliton-ode}. 
Reduction of the momentum and mass conservation laws is given by $(X +\mu Y-\nu T)|_{u=U(\xi)}=C=\const$, 
where $(T,X,Y)$ are expressions \eqref{conslaw1} and \eqref{conslaw2}. 
This gives two, functionally independent first integrals of this ODE. 
When the asymptotic conditions $U,U',U''\to 0$ as $|\xi|\to\infty$ are imposed, 
the first integrals yield the separable ODE
\begin{equation}\label{mgKP-U-energy}
U'{}^2  = (\nu-\sigma_2\mu^2) U^2 -\tfrac{1}{(q+1)(2q+1)}\sigma_1 U^{2q+2} - \tfrac{2}{(q+1)(q+2)}(a+b)\mu U^{q+2} .
\end{equation}

There are several different types of solutions to the ODE \eqref{mgKP-U-energy}. 
To obtain the line-soliton solutions,
we need to derive necessary and sufficient conditions
on coefficients in the ODE 
so that $|U(\xi)|$ has a single peak $|U_*|$ at a finite value $\xi=\xi_0$
and decays exponentially to $0$ for large $|\xi|$. 
This is readily carried out by applying a standard energy method
in which we write the ODE in the nonlinear oscillator form
\begin{equation}\label{ODE-oscil-eqn}
 U'{}^2 + V(U) =0
\end{equation}
where $U'{}^2$ is viewed as the kinetic energy term,
and where 
\begin{equation}\label{ODE-V}  
V(U) = -A U^2 +B U^{2q+2} + 2C U^{q+2}
\end{equation}
is viewed as the potential energy term,
with
\begin{align}
A & =\nu -\sigma_2\mu^2 , 
\label{A}\\
B & =\tfrac{1}{(q+1)(2q+1)}\sigma_1 , 
\label{B}\\
C &= \tfrac{1}{(q+1)(q+2)}(a+b)\mu . 
\label{C}
\end{align}
For later, we let
\begin{equation}\label{Delta}
\Delta=C^2+AB
=  \frac{\sigma_1(q+1)(q+2)^2(\nu-\mu^2 \sigma_2) +(2q+1)(a+b)^2\mu^2}{(2q+1)(q+1)^2(q+2)^2} .
\end{equation}
A straightforward analysis of the oscillator equation \eqref{ODE-oscil-eqn}
for $U\to0$ and $U\to U_*\neq0$ gives the following conditions. 

\begin{prop}\label{prop:Vconditions}
A line-soliton solution $U(\xi)$,
with a single peak $|U_*|$ at $\xi=\xi_0$
and with exponential decay $U\to 0$ for $|\xi|\to\infty$, 
arises iff the potential \eqref{ODE-V}
has the properties:
\begin{enumerate}[label=\arabic*.]
\item
$V(U) \simeq -A U^2$ for $|U|\ll 1$, with $A>0$,
so that $|\xi|\simeq O(\ln|U|)$ as $U\to0$;\\
\item
$U=U_*\neq 0$ is a root of $V(U)$ with $V'(U_*)>0$,
so that $\xi\simeq O(1)$ as $U\to U_*$.
\end{enumerate}
If property 2 is changed to $V(U_*)=V'(U_*)=0$,
so that $|\xi|\simeq O(\ln|U-U_*|)$ as $U\to U_*\neq0$, 
then $U(\xi)$ will be a line-shock solution. 
\end{prop}

Property 1 is established by noting:
$U'{}^2 \simeq AU^2$
$\Longleftrightarrow$
$\ln|U|\simeq \pm\sqrt{A}\xi$
$\Longleftrightarrow$
$|U|\simeq \exp(\pm\sqrt{A}\xi)$.
Similarly, property 2 is established by noting:
$U'{}^2 \simeq V'(U_*)(U_*-U)$
$\Longleftrightarrow$
$\sqrt{U_*-U} \simeq \tfrac{1}{2}V'(U_*)(\xi-\xi_0)$
$\Longleftrightarrow$
$U\simeq U_* +O((\xi-\xi_0))^2$. 

The resulting solution $U(\xi)$ will describe
a \emph{bright} line-soliton if $U_*>0$
or a \emph{dark} line-soliton if $U_*<0$,
and likewise in the limit of a line-shock. 

To apply Proposition~\ref{prop:Vconditions},
we first examine the roots of $V(U)$, 
with $A>0$. 
By factoring, we obtain
\begin{align}\label{V-factored}
V(U)= \tfrac{1}{B} U^2((BU^q +C)^2-\Delta) , 
\end{align}
whence the non-zero roots of $V(U)$ are determined by
$U_{\mathstrut *}^q =(\pm\sqrt{\Delta} -C)/B$, $\Delta\geq 0$. 
The nature of these roots depends on whether the power $q=\tfrac{1}{2}p$
is an odd or even integer when $p$ is an even integer, 
or is a half integer when $p$ is an odd integer. 

If $q$ is an odd integer 
then we have
\begin{align}\label{Vroots-q-oddcase}
U_{\mathstrut *}=\big((\pm\sqrt{\Delta} -C)/B\big)^{1/q} . 
\end{align}
Hence, 
$V(U)$ has $0,1,2$ roots when $\Delta <0$, $\Delta=0$, $\Delta>0$,
respectively.
The two properties in Proposition~\ref{prop:Vconditions}
for existence of a line-soliton solution
are satisfied only in the case $\Delta>0$, $A>0$. 

When $B<0$, both of the roots \eqref{Vroots-q-oddcase} have the same sign.
This implies that the line-soliton is either bright if the roots are positive,
or dark if the roots are negative.
The peak is given by the root closest to $0$.

In contrast, when $B>0$, the roots \eqref{Vroots-q-oddcase} have opposite signs,
and then there are pair of bright/dark line-solitons.
Their respective peaks are given by the positive root and the negative root. 

In the case $\Delta=0$, $A>0$ and $B<0$,
a line-shock solution arises instead,
because $V(U)$ has a repeated root.
The line-shock is either bright if the root is positive,
or dark if the root is negative.

If $q$ is an even integer 
then the roots have a different, more complicated form
which depends on the signs of $B$ and $C$, 
in addition to the $\Delta$.

When $B>0$, 
$V(U)$ has two roots with $A>0$ iff $\Delta >0$.
The roots are given by
\begin{equation}
U_{\mathstrut *}=\pm\big((\sqrt{\Delta} -C)/B\big)^{1/q} . 
\label{V2roots-q-evencase}
\end{equation}
In this situation,
Proposition~\ref{prop:Vconditions} shows that
there will be a symmetrical pair of bright/dark line-solitons. 

In contrast, when $B<0$,
$V(U)$ has four roots with $A>0$ for $\Delta >0$.
The roots are given by 
\begin{equation}
U_{\mathstrut *}=\big((\sqrt{\Delta} -C)/B\big)^{1/q},-\big((\sqrt{\Delta} -C)/B\big)^{1/q} , 
\label{V4roots-q-evencase}
\end{equation}
which comprise two symmetrical pairs.
Proposition~\ref{prop:Vconditions} again shows that
there will be a symmetrical pair of bright/dark line-solitons. 
Their peaks are given by the two roots closest to $0$. 

For $\Delta=0$ and $B<0$, 
$V(U)$ instead has two repeated roots with $A>0$ iff $C<0$.
The roots are a symmetrical pair given by 
\begin{equation}
U_{\mathstrut *}=\pm \big(-C/B\big)^{1/q} . 
\label{V2roots-alt-q-evencase}
\end{equation}
In this situation,
Proposition~\ref{prop:Vconditions} shows that 
a symmetrical pair of bright/dark line-shock solutions arises.

Finally, if $q$ is a half-integer
then the situation is the same as the odd integer case with the restriction
that $U_*>0$.
In particular,
when $B<0$, 
a line-soliton solution exists only for $\Delta>0$, $A>0$, $C>0$,
where $U_{\mathstrut *}$ is given by expression \eqref{Vroots-q-oddcase},
with the peak being the root closest to $0$.
When $B>0$, a line-soliton solution exists for $A>0$,
where the root is given by 
\begin{align}\label{Vroots-2q-oddcase}
U_{\mathstrut *}  =\big((\sqrt{\Delta} -C)/B\big)^{1/q} . 
\end{align}
In the case $\Delta=0$, $A>0$, $B<0$ and $C>0$, 
a line-shock solution exists. 

We will next obtain the explicit form of the preceding solutions
by integration of the separable ODE \eqref{ODE-oscil-eqn}
using the factorized form \eqref{V-factored} for $V(U)$. 

\begin{thm}\label{thm:solns}
All line-soliton and line-shock solutions $u=U(\xi)$ of the modified gKP equation \eqref{mgKP-scaled}
when the power $p=2q$ is a positive integer 
consist of:
\\
\indent (i) $q$ is an even integer \\
If $\sigma_1=\sgn(B)= 1$, then 
\begin{align}\label{bdsymm-solitonpair-sigma1ispos1}
U = \frac{\pm A^{1/q}}{(C+ \sqrt{C^2+AB}\cosh(q\sqrt{A}\xi))^{1/q}},
\quad
A>0
\end{align}
is a symmetrical pair of bright/dark line-solitons.
If $\sigma_1=\sgn(B)=-1$, then 
\begin{align}\label{bdsymm-solitonpair-sigma1isneg1}
U = \frac{\pm A^{1/q}}{(C+ \sqrt{C^2-A|B|}\cosh(q\sqrt{A}\xi))^{1/q}},
\quad
C>0,
\quad
C^2/|B|>A>0
\end{align}
is a symmetrical pair of bright/dark line-solitons; 
\begin{align}\label{bdsymm-shockpair}
U = \frac{\pm A^{1/q}}{(C(1+\exp(-q\sqrt{A}\xi)))^{1/q}},
\quad
C>0,
\quad
A=C^2/|B|
\end{align}
is symmetrical pair of bright/dark line-shocks. 
\\
\indent (ii) $q$ is an odd integer\\
If $\sigma_1=\sgn(B)=1$, then 
\begin{align}\label{bd-solitonpair}
U = \frac{A^{1/q}}{(C \pm \sqrt{C^2+AB}\cosh(q\sqrt{A}\xi))^{1/q}},
\quad
A>0
\end{align}
is a pair of bright/dark line-solitons.
They are symmetrical iff $C=0$.
If $\sigma_1=\sgn(B)=-1$, then 
\begin{align}\label{bd-soliton}
U = \frac{\sgn(C)A^{1/q}}{(|C|+ \sqrt{C^2-A|B|}\cosh(q\sqrt{A}\xi))^{1/q}},
\quad
C^2/|B|>A>0
\end{align}
is a line-soliton; 
\begin{align}\label{bd-shock}
U = \frac{\sgn(C)A^{1/q}}{(|C|(1+\exp(-q\sqrt{A}\xi)))^{1/q}},
\quad
A= C^2/|B|>0
\end{align}
is a line-shock. 
These solutions are bright or dark according to whether $\sgn(C)$ is $+1$ or $-1$.
\\
\indent (iii) $q$ is a half-integer\\
If $\sigma_1=\sgn(B)=1$, then 
\begin{align}\label{b-soliton-sigma1ispos1}
U = \frac{A^{1/q}}{(C + \sqrt{C^2+AB}\cosh(q\sqrt{A}\xi))^{1/q}},
\quad
A>0
\end{align}
is a line-soliton.
If $\sigma_1=\sgn(B)=-1$, then 
\begin{align}\label{b-soliton-sigma1isneg1}
U = \frac{A^{1/q}}{(C+ \sqrt{C^2-A|B|}\cosh(q\sqrt{A}\xi))^{1/q}},
\quad
C>0,
\quad
C^2/|B|>A>0
\end{align}
is a line-soliton; 
\begin{align}\label{b-shock}
U = \frac{A^{1/q}}{(C(1+\exp(-q\sqrt{A}\xi)))^{1/q}},
\quad
C>0,
\quad
A=C^2/|B|>0
\end{align}
is a line-shock. 
All of these solutions are bright. 
\end{thm}

We remark that Theorem~\ref{thm:solns} 
can be extended to situation when $p$ is rational number as follows:
case $(i)$ holds with $q$ being a rational number with an even numerator;
case $(ii)$ holds with $q$ a rational number with an odd numerator and an odd denominator;
case $(iii)$ holds with $q$ a rational number with an odd numerator and an even denominator.

We also remark that for the case $C=0$ the line solitons in Theorem~\ref{thm:solns} 
reduce to the ones found in \Ref{AncGanRec2018} for the gKP equation
with an arbitrary power $p>0$. 
In particular, in this case 
the pairs of bright/dark line-solitons 
\eqref{bdsymm-solitonpair-sigma1ispos1} and \eqref{bd-solitonpair} coincide
with the gKP line-solitons when $p$ is an even integer;
and the single bright soliton \eqref{b-soliton-sigma1ispos1}
corresponds to the gKP line-soliton when $p$ is an odd integer.

Consequently, hereafter we will restrict attention to the modified gKP case
where $C\neq 0$.

\subsection{Kinematical properties of line-solitons and line-shocks}

We will next discuss the main kinematical properties of the modified gKP line-soliton and line-shock solutions:
their speed and direction, width, and height.

With respect to the $x$ axis, 
the angle of the (tilted) line of motion of the line solutions
is given by 
\begin{equation}\label{angle}
\theta=\arctan\mu,
\end{equation}
while the speed of the line solution along this tilted line is given by
\begin{equation}\label{speed}
c=\nu/\sqrt{1+\mu^2},
\end{equation}
with the sign of $\nu$ specifying the direction of propagation. 
Note that the two parameters $(\mu,\nu)$ need to obey conditions 
corresponding to the conditions on $A,B,C$ in Theorem~\ref{thm:solns}.
The explicit form of these conditions will be presented in the next subsections.

The width of the line solutions is proportional to
\begin{equation}
w=1/(q\sqrt{A}) .
\end{equation}
Their heights/depths are given by
\begin{equation}
h = A^{1/q}/(|C|+ \sqrt{C^2-A|B|})^{1/q}
\end{equation}
for the single bright/dark line-solitons \eqref{bd-soliton};
\begin{equation}
h = A^{1/q}/(C+ \sqrt{C^2+\sigma_1 A|B|})^{1/q}
\end{equation}
for the bright line-solitons \eqref{b-soliton-sigma1ispos1}--\eqref{b-soliton-sigma1isneg1}
and for the symmetrical pairs of line-solitons \eqref{bdsymm-solitonpair-sigma1ispos1}--\eqref{bdsymm-solitonpair-sigma1isneg1};
\begin{equation}
h_\pm = A^{1/q}/|C \pm\sqrt{C^2+AB}|^{1/q}
\end{equation}  
for the non-symmetrical pair of line-solitons \eqref{bd-solitonpair};
\begin{equation}
h = (A/|C|)^{1/q}
\end{equation}  
for the bright/dark line-shocks \eqref{bd-shock} and \eqref{b-shock}; 
and 
\begin{equation}
h = (A/C)^{1/q}
\end{equation}  
for the symmetrical pair of line-shocks \eqref{bdsymm-shockpair}.

To discuss the profile shapes, 
we will examine the line-solitons first,
and the line-shocks last.

\subsection{Line-soliton profiles}

We will use the height and the width
to parameterize the line-soliton profiles,
as well as their speed and the direction angle. 

\begin{thm}
In terms of height $h>0$ and width $w>0$, the profiles of
the line-solitons \eqref{bd-soliton}, \eqref{b-soliton-sigma1ispos1}, \eqref{b-soliton-sigma1isneg1},
the symmetrical pairs of line-solitons \eqref{bdsymm-solitonpair-sigma1ispos1}--\eqref{bdsymm-solitonpair-sigma1isneg1},
and the non-symmetrical pair of line-solitons \eqref{bd-solitonpair}
have the form 
\begin{equation}\label{soliton-profile}
|U|= \frac{h}{( (1 +\sigma_1(w \h^q/l)^2)\cosh(\xi/(2w))^2 -\sigma_1 (w \h^q/l)^2 )^{1/q}} ,
\end{equation}
where 
\begin{equation}\label{lsq}
l^2 =(2q+1)(q+1)/q^2 . 
\end{equation}
In the focussing case $\sigma_1=1$,
there are no conditions on $h$ and $w$,
while in the defocussing case $\sigma_1=-1$,
they must obey the condition 
\begin{equation}
  w \h^q < l
  \quad (\sigma_1=-1)
\end{equation}
so that $U$ is non-singular. 
In both cases, the direction angle and the speed are given by 
\begin{equation}\label{solitary-angl}
|\theta| = \arctan\left(\frac{m^2|1-\sigma_1 w^2 \h^{2q}/l^2|}{2|a+b|w^2 \h^q}\right)
\end{equation}
and
\begin{equation}\label{solitary-speed}
c = \frac{\sigma_2 q^2 m^4 (1 -\sigma_1 w^2 \h^{2q}/l^2)^2 +4(a+b)^2 w^2 \h^{2q}}{2 (a+b) q^2 w^2 \h^q \sqrt{m^4(1 -\sigma_1 w^2 \h^{2q}/l^2)^2+4(a+b)^2 w^4 \h^{2q}}} ,
\end{equation}
where
\begin{equation}\label{msq}
m^2 = (q+1)(q+2)/q^2 . 
\end{equation}  
\end{thm}
(Information about the sign of $\theta$ will be given in section~\ref{sec:kinematics}.)

\begin{figure}[h]
\centering
\includegraphics[width=.32\textwidth,trim=2cm 18cm 8cm 4cm,clip]{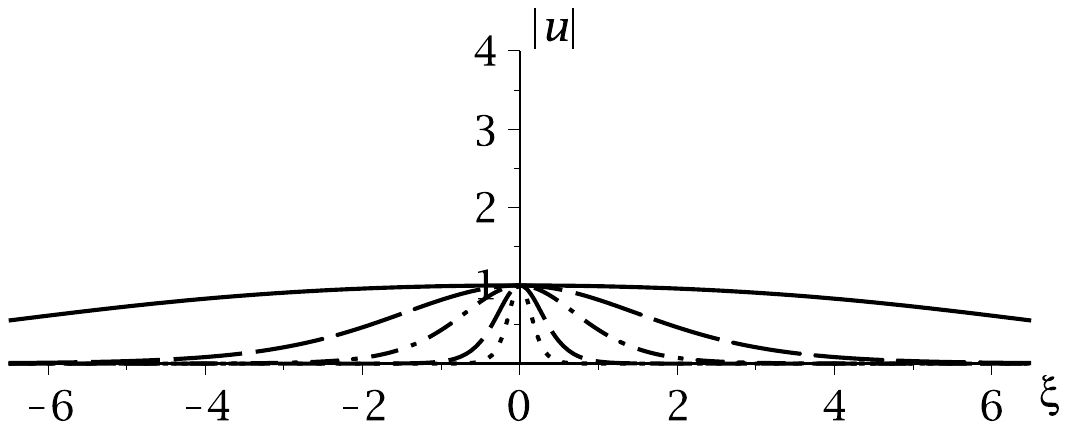}
\includegraphics[width=.32\textwidth,trim=2cm 18cm 8cm 4cm,clip]{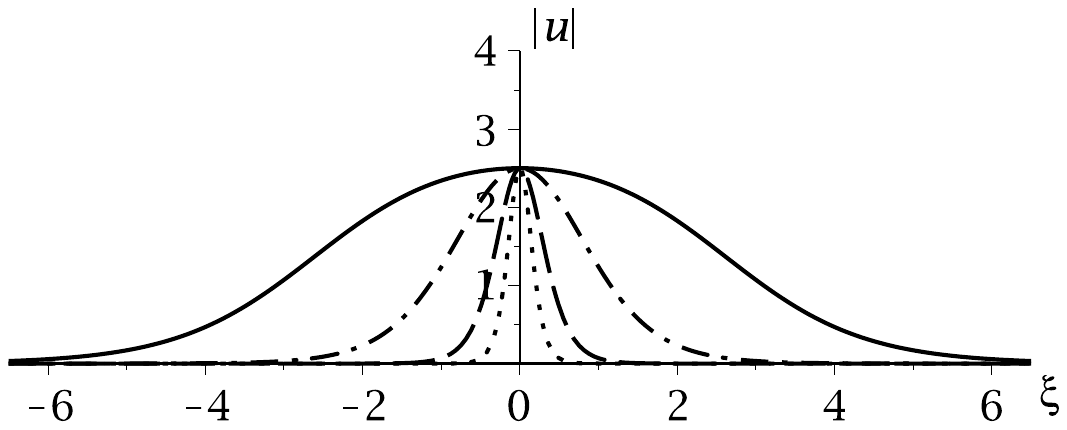}
\includegraphics[width=.32\textwidth,trim=2cm 18cm 8cm 4cm,clip]{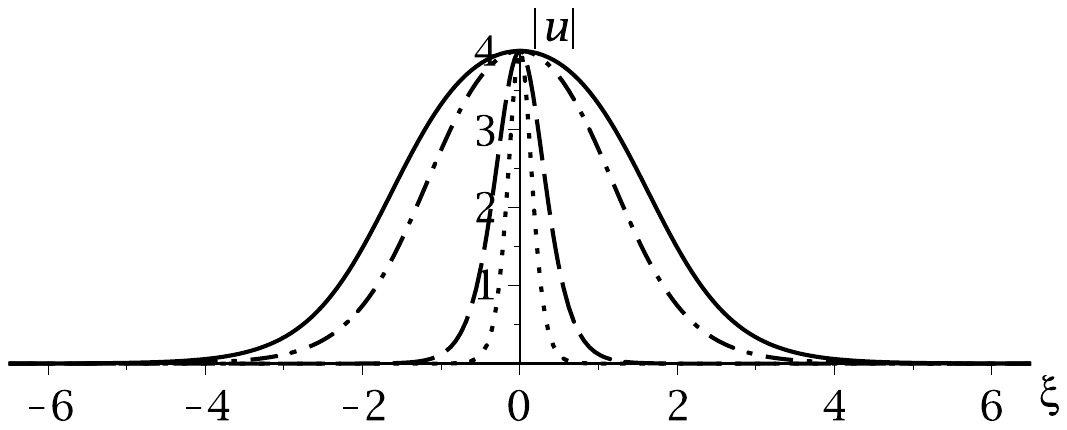}
\caption{Line-soliton profile in the defocussing case.
$q=1$; 
$h=$
$1$ (left), $2.5$ (middle), $4$ (right);
$w=$ 
$\tfrac{1}{10}$ (dots), $\tfrac{1}{5}$ (dashes), $\tfrac{1}{2}$ (dot-dashes), $1$ (long-dashes), $w\approx 0.90 w_{\text{max}}$ (solid).}
\label{fig:qis1-sigma1isneg1-profile}
\end{figure}

\begin{figure}[h]
\centering
\includegraphics[width=.32\textwidth,trim=2cm 18cm 8cm 4cm,clip]{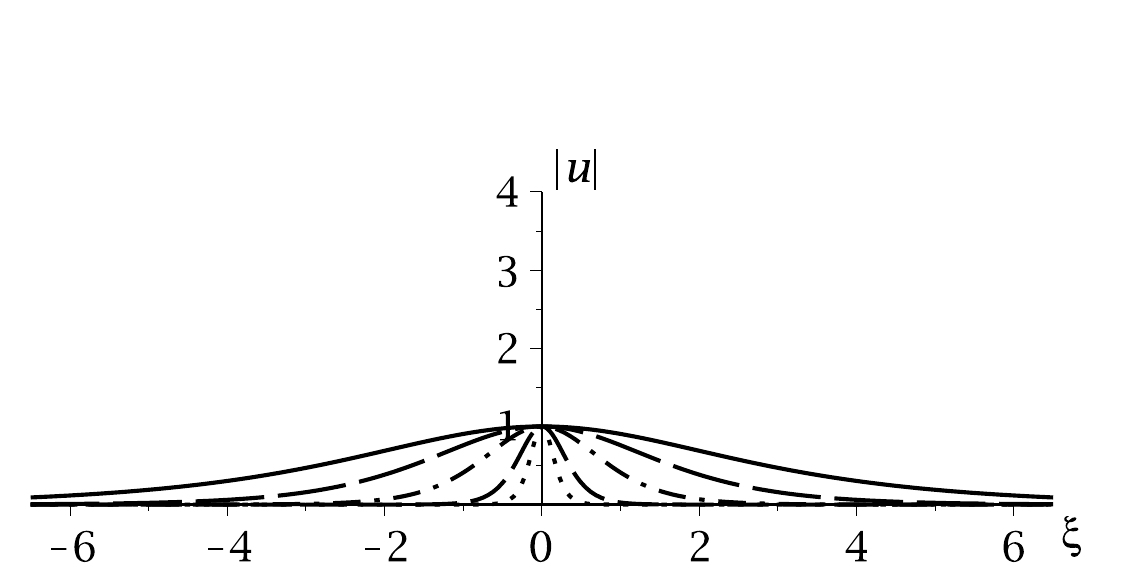}
\includegraphics[width=.32\textwidth,trim=2cm 18cm 8cm 4cm,clip]{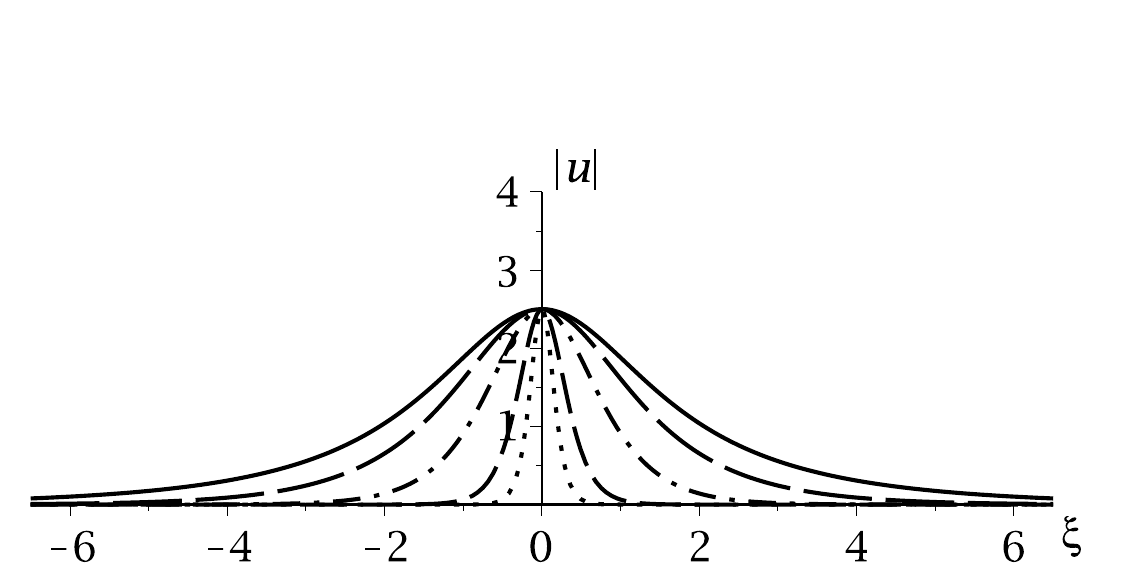}
\includegraphics[width=.32\textwidth,trim=2cm 18cm 8cm 4cm,clip]{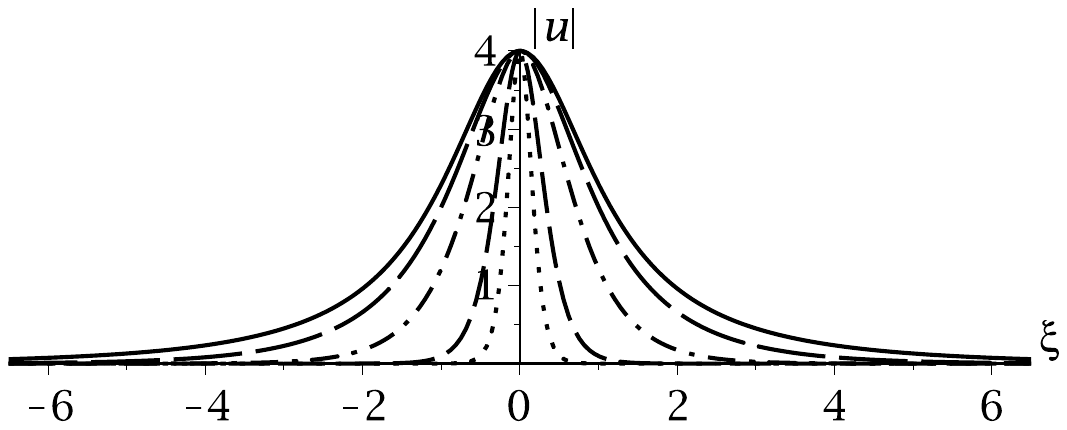}  
\caption{Line-soliton profile in the focussing case. 
$q=1$; 
$h=$
$1$ (left), $2.5$ (middle), $4$ (right);
$w=$
$\tfrac{1}{10}$ (dots), $\tfrac{1}{5}$ (dashes), $\tfrac{1}{2}$ (dot-dashes), $1$ (long-dashes), $2$ (solid).}
\label{fig:qis1-sigma1is1-profile}
\end{figure}

The profile \eqref{soliton-profile} for fixed $w$ and $h$
differs in the cases $\sigma_1=1$ and $\sigma_1=-1$
but is qualitatively similar for all values of $q$. 
Plots are shown in
Figs.~\ref{fig:qis1-sigma1isneg1-profile} and~\ref{fig:qis1-sigma1is1-profile}. 

\subsection{Line-shock profiles}

We will use the height and the width
to parameterize the line-shock profile,
as well as their speed and the direction angle.
In contrast to line-solitons,
there is no restriction on $h$ and $w$ for line-shocks. 

\begin{thm}
In terms of height $h>0$ and width $w>0$, 
the profiles of the symmetrical pair of line-shocks \eqref{bdsymm-shockpair},
and the single line-shocks \eqref{bd-shock} and \eqref{b-shock}
have the form 
\begin{equation}\label{shock-profile}
|U|= \frac{h}{(1+ \exp(-\xi/w))^{1/q}} ,
\quad
\sigma_1=-1 . 
\end{equation}
The direction angle and the speed are given by 
\begin{equation}\label{shock-angl}
\theta =\arctan\left(\frac{m^2 \h^{q}}{l^2(a+b)}\right)
\end{equation}
and
\begin{equation}\label{shock-speed}
c = \frac{k^2l^4 +\sigma_2 q^2 m^4 w^2 \h^{2q}}{(a+b) q^2l^2 w^2\sqrt{(a+b)^2 l^4 +m^4\h^2}} , 
\end{equation}
where $l^2$ and $m^2$ are given by expressions \eqref{lsq} and \eqref{msq}.
\end{thm}

\begin{figure}[h]
\centering
\includegraphics[width=.32\textwidth,trim=2cm 17cm 8cm 4cm,clip]{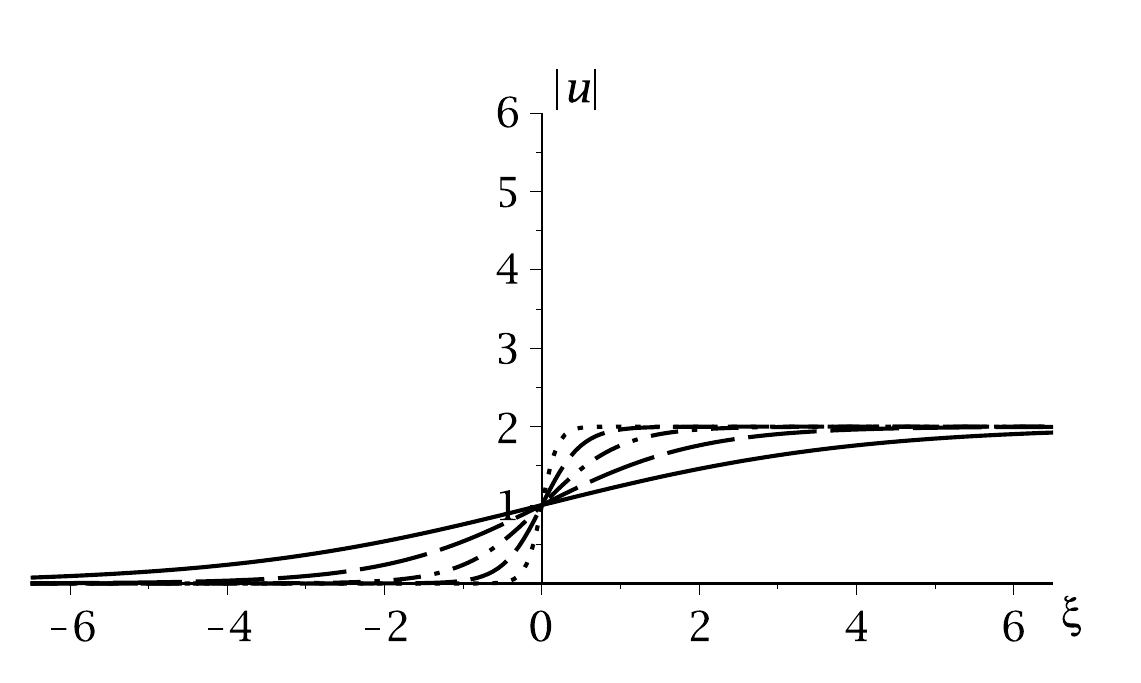}
\includegraphics[width=.32\textwidth,trim=2cm 17cm 8cm 4cm,clip]{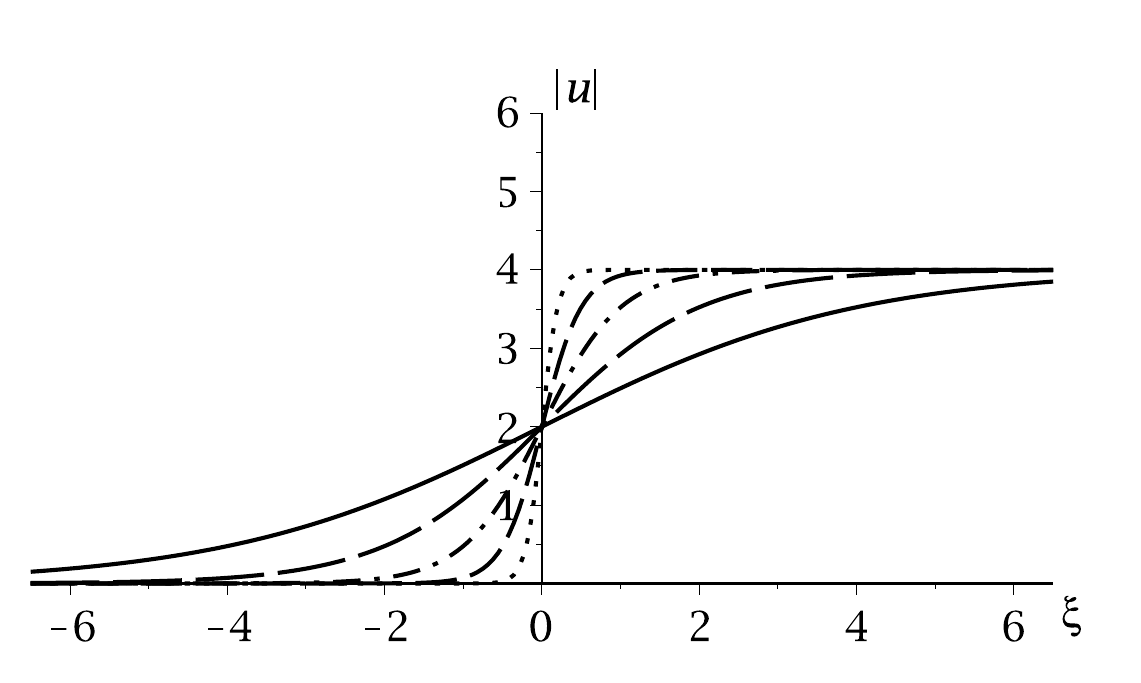}
\includegraphics[width=.32\textwidth,trim=2cm 17cm 8cm 4cm,clip]{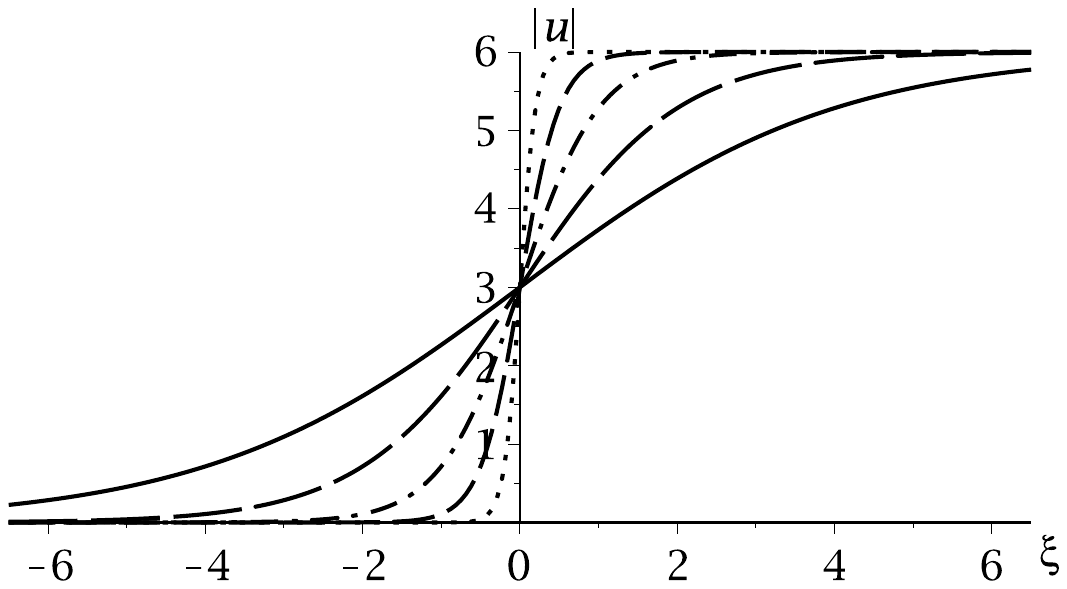}
\caption{Line-shock profile. 
$q=1$; 
$h=$
$2$ (left), $4$ (middle), $6$ (right);
$w=$ 
$\tfrac{1}{10}$ (dots), $\tfrac{1}{4}$ (dashes), $\tfrac{1}{2}$ (dot-dashes), $1$ (long-dashes), $2$ (solid).}
\label{fig:shock-qis1-profile}
\end{figure}

The profile \eqref{shock-profile} for fixed $w$ and $h$
is qualitatively similar for all values of $q$. 
Plots are shown in Fig.~\ref{fig:shock-qis1-profile}. 

\section{Kinematics features in terms of the speed and the angle}\label{sec:kinematics}

We will examine in detail 
the properties of the modified gKP line solutions in Theorem~\ref{thm:solns}
by using a physical parameterization given by their speed $c$ and direction angle $\theta$.

\subsection{Kinematics of line-solitons}

In every line-soliton solution, 
the parameters $\mu$ and $\nu$ must satisfy the condition
\begin{equation}\label{param-Aispos}
\nu>\sigma_2\mu^2
\end{equation}
corresponding to $A>0$. 
The additional conditions $C>0$ and $C^2>A|B|$
required for existence of some of the line-soliton solutions
correspond to
\begin{equation}\label{param-Cispos}
(a+b)\mu >0
\end{equation}
and
\begin{equation}\label{param-Deltaispos}
\nu < (\sigma_2 +\tfrac{l^2}{m^4}(a+b)^2)\mu^2
\end{equation}  
respectively. 
These conditions can be expressed in terms of the speed and the angle, 
via inverting the relations \eqref{angle}--\eqref{speed}.

\begin{prop}\label{prop:soliton-kinematic-conds}
The speed $c$ and the direction angle $\theta$ of all line-soliton solutions
of the modified gKP equation \eqref{mgKP-scaled}
obey the kinematic condition
\begin{equation}\label{speed-angle-Aispos}
c>\sigma_2 \sin^2 \theta/\cos\theta,
\quad
-\tfrac{1}{2}\pi < \theta \leq \tfrac{1}{2}\pi . 
\end{equation}
No further conditions are required by
the symmetrical pair of bright/dark line-solitons \eqref{bdsymm-solitonpair-sigma1ispos1},
the non-symmetrical pair of bright/dark line-solitons \eqref{bd-solitonpair},
and the bright line-soliton \eqref{b-soliton-sigma1ispos1},
all of which have $\sigma_1=1$. 
The single bright/dark line-soliton \eqref{bd-soliton} 
requires an additional kinematic condition
\begin{equation}\label{speed-angle-Deltaispos}
c < (\sigma_2+ \k^2) \sin^2 \theta/\cos\theta ,
\quad
\sigma_1=-1 
\end{equation}
where
\begin{equation}\label{k}
\k = \tfrac{\sqrt{2q+1}}{\sqrt{q+1}(q+2)}(a+b) . 
\end{equation}
This condition \eqref{speed-angle-Deltaispos}
and another kinematic condition
\begin{equation}\label{speed-angle-Cispos}
\sgn\;\theta = \sgn\;\k
\end{equation}
are required by
both the other bright line-soliton \eqref{b-soliton-sigma1isneg1}
and the other symmetrical pair of bright/dark line-solitons \eqref{bdsymm-solitonpair-sigma1isneg1}. 
\end{prop}

Since these kinematic conditions depend on both $\sigma_1$ and $\sigma_2$,
we will organize the subsequent discussion into four cases: 
focussing and defocussing cases, $\sigma_1 =+1,-1$; 
normal and sign-changing dispersion cases, $\sigma_2 =+1,-1$.
The KP-like cases are $\sigma_1=1$, $\sigma_2=\pm1$;
the mKP-like cases are $\sigma_1=-\sigma_2=1$
and $\sigma_1=-\sigma_2=-1$. 

\subsubsection{Focussing with normal dispersion}

When $\sigma_1=\sigma_2=1$,
the kinematical parameters $(\nu,\mu)$ satisfy
$0\leq \mu^2<\nu<\infty$.
From Proposition~\ref{prop:soliton-kinematic-conds}, 
there is a minimum speed
\begin{equation}\label{cmin-normaldispersion}
c_{\text{min}}(\theta) = \sin^2 \theta/\cos\theta,
\end{equation}
which is positive for all directions $-\frac{\pi}{2}<\theta<\frac{\pi}{2}$,
while the maximum speed is unbounded.
Interestingly, as the direction becomes more transverse,
the minimum speed is higher. 

\begin{figure}[h]
\centering
\includegraphics[width=0.45\textwidth]{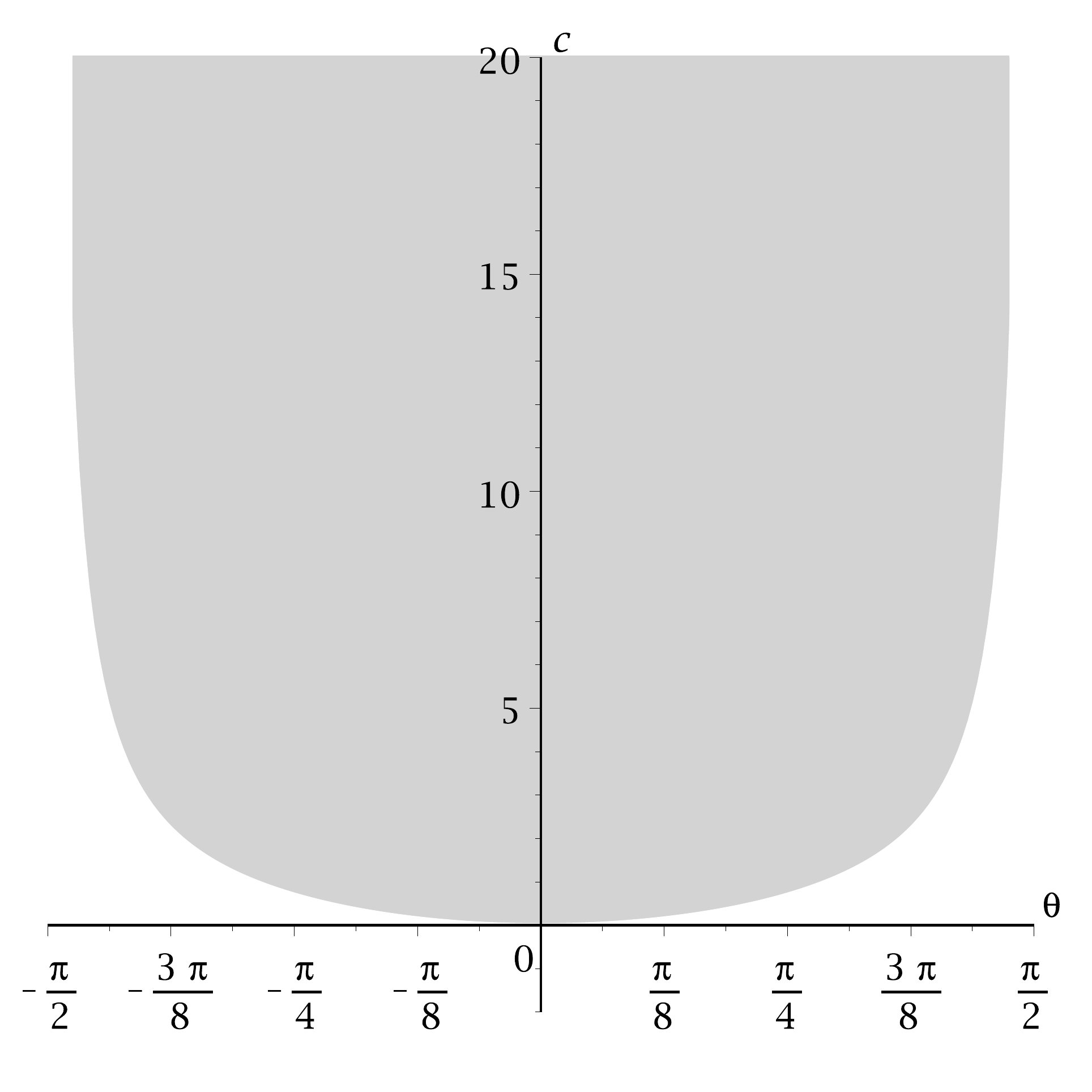}
\caption{Kinematically allowed region in $(c,\theta)$ for line-solitons
in the case of focussing with normal dispersion.}
\label{fig:sigma1is1-sigma2is1-kinregion}
\end{figure}

For a fixed speed $c>c_{\text{min}}(\theta)$,
the direction angle has the range
$-\vartheta(c) < \theta < \vartheta(c)$,
where
\begin{equation}\label{theta_c_rel}
\vartheta(c) =
\arctan\textstyle{ \sqrt{\tfrac{1}{2}c(\sqrt{c^2+4} +c}) }
\end{equation}
is the angle determined by $c=c_{\text{min}}(\theta)$. 
The kinematically allowed region in $(c,\theta)$ is plotted 
in Fig.~\ref{fig:sigma1is1-sigma2is1-kinregion}. 
Note that this region is independent of the nonlinearity power $p$. 

The modified gKP line-solitons for $\theta\neq 0$ can be expressed 
in terms of $c$ and $\theta$ as
\begin{equation}\label{u-solitary-focusnormaldisper}
u=\dfrac{s((q+1)(2q+1))^\frac{1}{2q}((c/c_{\text{min}}(\theta)-1)\tan|\theta|)^\frac{1}{q}}
{\Big( \tilde s\k +\sqrt{\k^2-1+c/c_{\text{min}}(\theta)}\cosh\big(q\sqrt{c/c_{\text{min}}(\theta)-1}\tan|\theta|\;(x+\tan\theta\;y-c\sec\theta\;t)\big) \Big)^\frac{1}{q}}
\end{equation}
where $s$ and $\tilde s$ are signs for the 3 types of line-solitons
shown in Table~\ref{table:focus-signs}. 

\begin{table}[h!]
\caption{Signs in the line-solitons: focussing case}
\label{table:focus-signs}
\centering
\begin{tabular}{c|c||c|c|c}
\hline
Type
& $q$
& $\tilde s$
& $s$
& $\sgn\,\theta$
\\
\hline
\hline
bright/dark symmetrical pair
&
even
&
$\sgn\,\theta$
&
$\pm 1$
&
$\gtrless0$
\\
\hline
bright/dark non-symmetrical pair
&
odd
&
$\pm\sgn\,\theta$
&
$\pm1$
&
$\gtrless0$
\\
\hline
bright
&
half-integer
&
$\sgn\,\theta$
&
$1$
&
$\gtrless0$
\\
\hline
\end{tabular}
\end{table}

The width and height are given by 
\begin{align}
w & = \frac{1}{q\sqrt{c/c_{\text{min}}(\theta)-1}\tan|\theta|} ,
\label{w-focusnormaldisper}
\\
h & =\big((q+1)(2q+1)\big)^\frac{1}{2q}\big(\big|\sqrt{c/c_{\text{min}}(\theta) +\k^2 -1} -\tilde s\k\big| \tan|\theta| \big)^\frac{1}{q} .
\label{h-focusnormaldisper}
\end{align}
At a fixed speed $c>c_{\text{min}}$,
the width is symmetrical under $\theta\to -\theta$
but the height is asymmetrical.

At a fixed direction angle $\theta$ with $|\theta|<\vartheta(c)$,
the width decreases and the height increases
as the speed $c$ increases and as the nonlinearity power $q$ increases.
In the limiting case of the minimum speed, 
the line-soliton flattens to $u=0$
when the angle has the sign given by $\tilde s= \sgn\,\k$.
An interesting question that we will explore elsewhere is what the limit of $u$
looks like when the angle has the opposite sign. 

\subsubsection{Defocussing with normal dispersion}

\begin{figure}[h]
\centering
\includegraphics[width=0.45\textwidth,trim=2cm 15cm 8cm 2cm,clip]{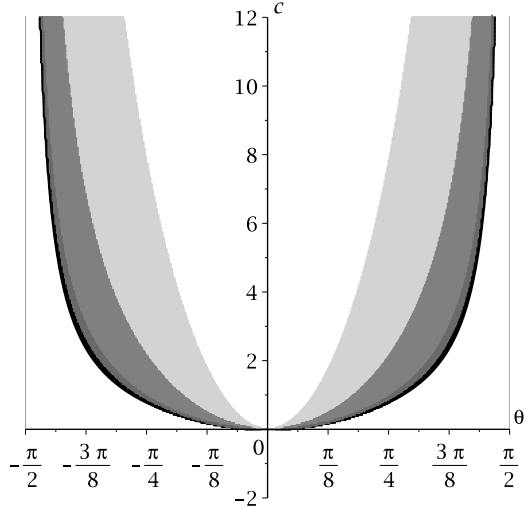}
\caption{Kinematically allowed region in $(c,\theta)$ for line-solitons
in the case of defocussing with normal dispersion. 
$\k^2=$ 
$\tfrac{1}{8}$ (black), 
$\tfrac{1}{3}$ (dark grey), 
$2$ (grey), 
$10$ (light grey); 
lighter regions overlap with all darker regions. }
\label{fig:sigma1isneg1-sigma2is1-kinregion}
\end{figure}

When $-\sigma_1=\sigma_2=1$,
the kinematical parameters $(\nu,\mu)$ satisfy
$\mu^2<\nu < (\k^2+1)\mu^2$. 
From Proposition~\ref{prop:soliton-kinematic-conds}, 
the minimum speed is the same as in the focussing case,
but the maximum speed is finite
\begin{equation}\label{cmax-normaldisper}
c_{\text{max}}(\theta) = \big(1 + \k^2\big)\sin^2\theta/\cos\theta,
\end{equation}  
which depends on $\k$. 
As a consequence, for a fixed speed $c_{\text{min}}(\theta)<c<c_{\text{max}}(\theta)$,
the angular range of the direction is
$\vartheta(c/(1+\k^2)) < |\theta| < \vartheta(c)$, 
where $\vartheta$ is the angle \eqref{theta_c_rel}.
The kinematically allowed region in $(c,\theta)$ is plotted 
in Fig.~\ref{fig:sigma1isneg1-sigma2is1-kinregion}. 
Note that the mKP case is given by $\k=\tfrac{1}{\sqrt{3}}$. 

The modified gKP line-solitons for $\theta\neq 0$ can be expressed 
in terms of $c$ and $\theta$ as
\begin{equation}\label{u-solitary-defocusnormaldisper}
u=\dfrac{s((q+1)(2q+1))^\frac{1}{2q}((c/c_{\text{min}}(\theta)-1)\tan|\theta|)^\frac{1}{q}}
{\Big( |k| +\sqrt{(\k^2+1)(1-c/c_{\text{max}}(\theta))}\cosh\big(q\sqrt{c/c_{\text{min}}(\theta)-1}\tan|\theta|\;(x+\tan\theta\;y-c\sec\theta\;t)\big) \Big)^\frac{1}{q}}
\end{equation}
where $s$ is a sign for the 3 types of line-solitons
shown in Table~\ref{table:defocus-signs}. 

\begin{table}[h!]
\caption{Signs in the line-solitons: defocussing case}
\label{table:defocus-signs}
\centering
\begin{tabular}{c|c||c|c}
\hline
Type
& $q$
& $s$
& $\sgn\,\theta$
\\
\hline
\hline
bright/dark symmetrical pair
&
even
&
$\pm 1$
&
$\sgn\,\k$
\\
\hline
bright/dark 
&
odd
&
$\sgn(\k\theta)$
&
$\gtrless0$
\\
\hline
bright
&
half-integer
&
$1$
&
$\sgn\,\k$
\\
\hline
\end{tabular}
\end{table}

The width has the same expression \eqref{w-focusnormaldisper} as in the previous case,
but the height is given by
\begin{equation}\label{h-defocusnormaldisper}
h=\big((q+1)(2q+1)\big)^\frac{1}{2q}\Big( \big|\sqrt{(\k^2+1)(1-c/c_{\text{max}}(\theta))}-|\k|\big| \tan|\theta| \Big)^\frac{1}{q} . 
\end{equation}
At a fixed speed, 
both the width and the height are symmetrical under $\theta\to -\theta$. 

At a fixed direction angle $\theta$,
with $\vartheta(c/(1+\k^2))<|\theta|<\vartheta(c)$,
the width decreases and the height increases
as the speed $c$ increases to $c_{\text{max}}$
and as the nonlinearity power $q$ increases.
In the limiting case of the maximum speed,
$u$ approaches a step-like function whose width increases logarithmically with $c_{\text{max}}-c$. 
In the opposite limiting case of the minimum speed, 
$u$ flattens to $0$. 

\subsubsection{Focussing with sign-changing dispersion}

When $\sigma_1=-\sigma_2=1$,
the kinematical parameters $(\nu,\mu)$ satisfy $-\mu^2<\nu < \infty$.
From Proposition~\ref{prop:soliton-kinematic-conds}, 
the maximum speed is unbounded while the minimum speed is negative,
and so the line-soliton can move forward or backward, or remain stationary, 
relative to the $x$-direction. 

\begin{figure}[h]
\centering
\includegraphics[width=0.45\textwidth]{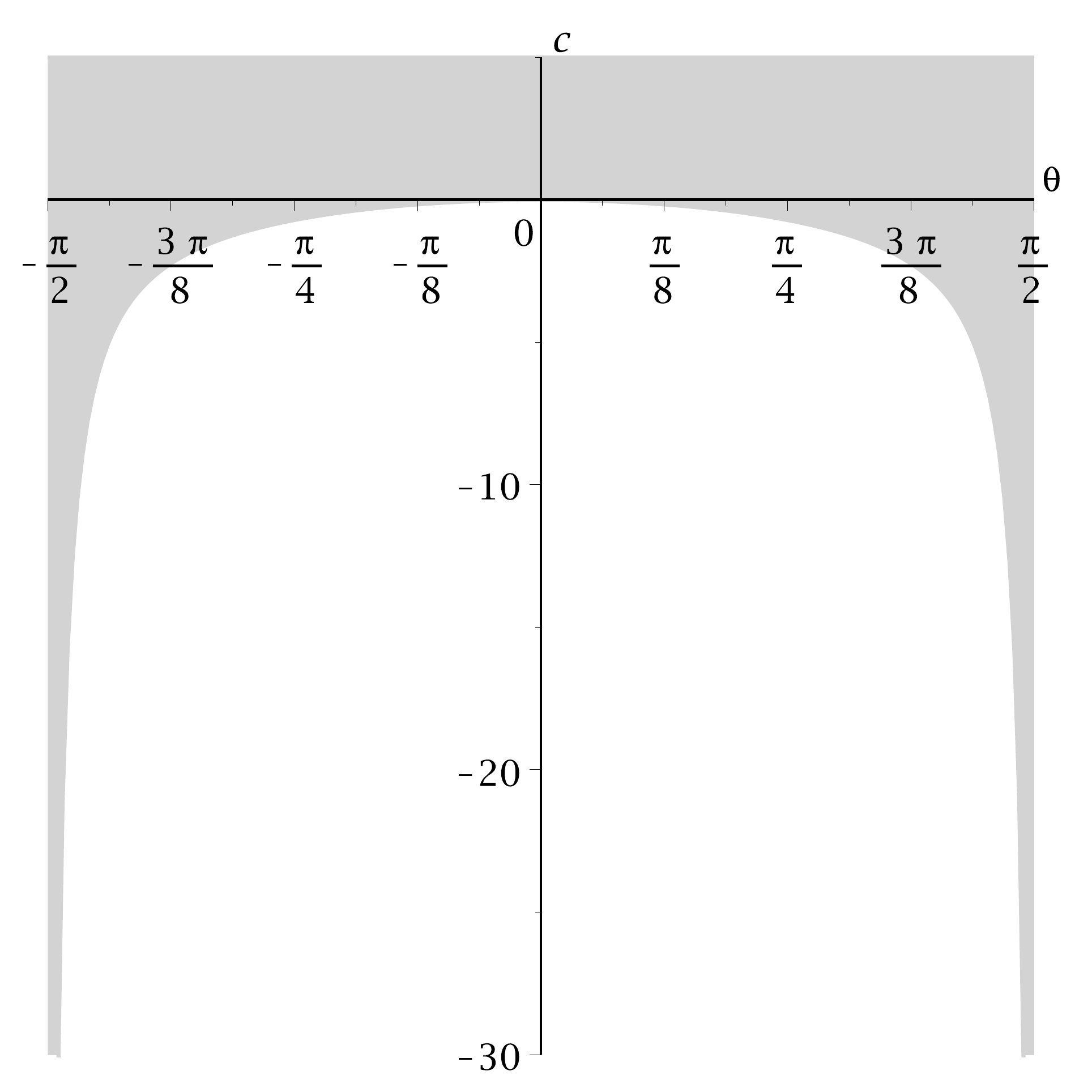}
\caption{Kinematically allowed region in $(c,\theta)$ for line-solitons
in the case of focussing with sign-changing dispersion.}
\label{fig:sigma1is1-sigma2isneg1-kinregion}
\end{figure}

The minimum speed is given by 
\begin{equation}\label{cmin-focusnegativedisper}
c_{\text{min}}(\theta) = -\sin^2\theta/\cos\theta . 
\end{equation}
For a fixed negative speed, 
the direction angle has the range
$\vartheta(|c|) < |\theta| < \tfrac{1}{2}\pi$, 
where $\vartheta$ is the angle \eqref{theta_c_rel}.
For a fixed non-negative speed,
the direction angle has the range
$0\leq |\theta| < \tfrac{1}{2}\pi$. 
The kinematically allowed region in $(c,\theta)$ is plotted 
in Fig.~\ref{fig:sigma1is1-sigma2isneg1-kinregion}. 
Note that this region is independent of the nonlinearity power $p$ 
and hence coincides with the kinematically allowed region for the mKP equation. 

The modified gKP line-solitons for $\theta\neq 0$ can be expressed as
\begin{equation}\label{u-solitary-focusnegdisper}
u=\dfrac{s((q+1)(2q+1))^\frac{1}{2q}((1+c/|c_{\text{min}}(\theta)|)\tan|\theta|)^\frac{1}{q}}
{\Big(\tilde s\k +\sqrt{\k^2+1+c/|c_{\text{min}}|}\cosh\big(q\sqrt{1+c/|c_{\text{min}}(\theta)|}\tan|\theta|(x+\tan\theta\;y-c\sec\theta\;t)\big) \Big)^\frac{1}{q}}
\end{equation}
where the signs $s$ and $\tilde s$ are given in Table~\ref{table:focus-signs}. 
This expression \eqref{u-solitary-focusnegdisper} differs from
the line-soliton \eqref{u-solitary-focusnormaldisper} in normal dispersion case 
by the change in sign of the minimum speed \eqref{cmin-focusnegativedisper}. 
Similarly, the width and height are given by
\begin{align}
w & = \frac{1}{q\sqrt{1+c/|c_{\text{min}}(\theta)|}\tan|\theta|} , 
\label{w-focusnegdisper}
\\
h & =\big((q+1)(2q+1)\big)^\frac{1}{2q}\big( \big|\sqrt{\k^2 +1+c/|c_{\text{min}}(\theta)|}-\tilde s\k\big| \tan|\theta| \big)^\frac{1}{q} . 
\label{h-focusnegdisper}
\end{align}
At a fixed speed, 
the width is symmetrical under $\theta\to -\theta$
while the height is asymmetrical.

At a fixed direction angle $\theta$,
with $\vartheta(c/(1+\k^2))<|\theta|<\vartheta(c)$,
the width decreases and the height increases
as the speed $c$ increases and as the nonlinearity power $q$ increases.
The limiting case of the minimum speed is similar to what occurs
for focussing with normal dispersion. 

\subsubsection{Defocussing with sign-changing dispersion}

When $-\sigma_1=-\sigma_2=1$,
the kinematical parameters $(\nu,\mu)$ satisfy $-\mu^2<\nu < (\k^2-1)\mu^2$ and $\mu >0$. 
From Proposition~\ref{prop:soliton-kinematic-conds}, 
the minimum speed is the same as in the previous case, 
while the maximum speed is given by 
\begin{equation}\label{cmax-defocusnegdisper}
c_{\text{max}} = ( \k^2-1)\sin^2\theta/\cos\theta, 
\end{equation}
which is positive, negative, or zero, depending on whether
$\k$ is larger, smaller, or equal to $1$. 

\begin{figure}[h]
\centering
\includegraphics[width=0.45\textwidth,trim=2cm 15cm 8cm 2cm,clip]{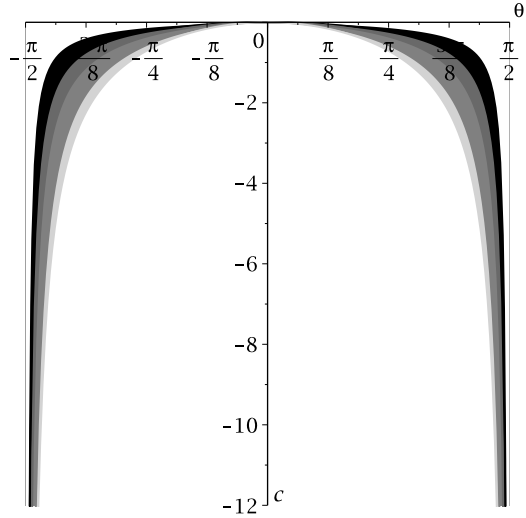}
\caption{Kinematically allowed region in $(c,\theta)$ for line-solitons
in the case of defocussing with sign-changing dispersion. 
$\k^2=$ 
$\tfrac{1}{5}$ (black), 
$\tfrac{1}{2}$ (dark grey), 
$\tfrac{2}{3}$ (grey), 
$\tfrac{5}{6}$ (light grey);
lighter regions overlap with all darker regions. 
}
\label{fig:sigma1isneg1-sigma2isneg1-kinregion-ksqlessthan1}
\end{figure}

\begin{figure}[h]
\centering
\includegraphics[width=0.45\textwidth,trim=2cm 15cm 8cm 2cm,clip]{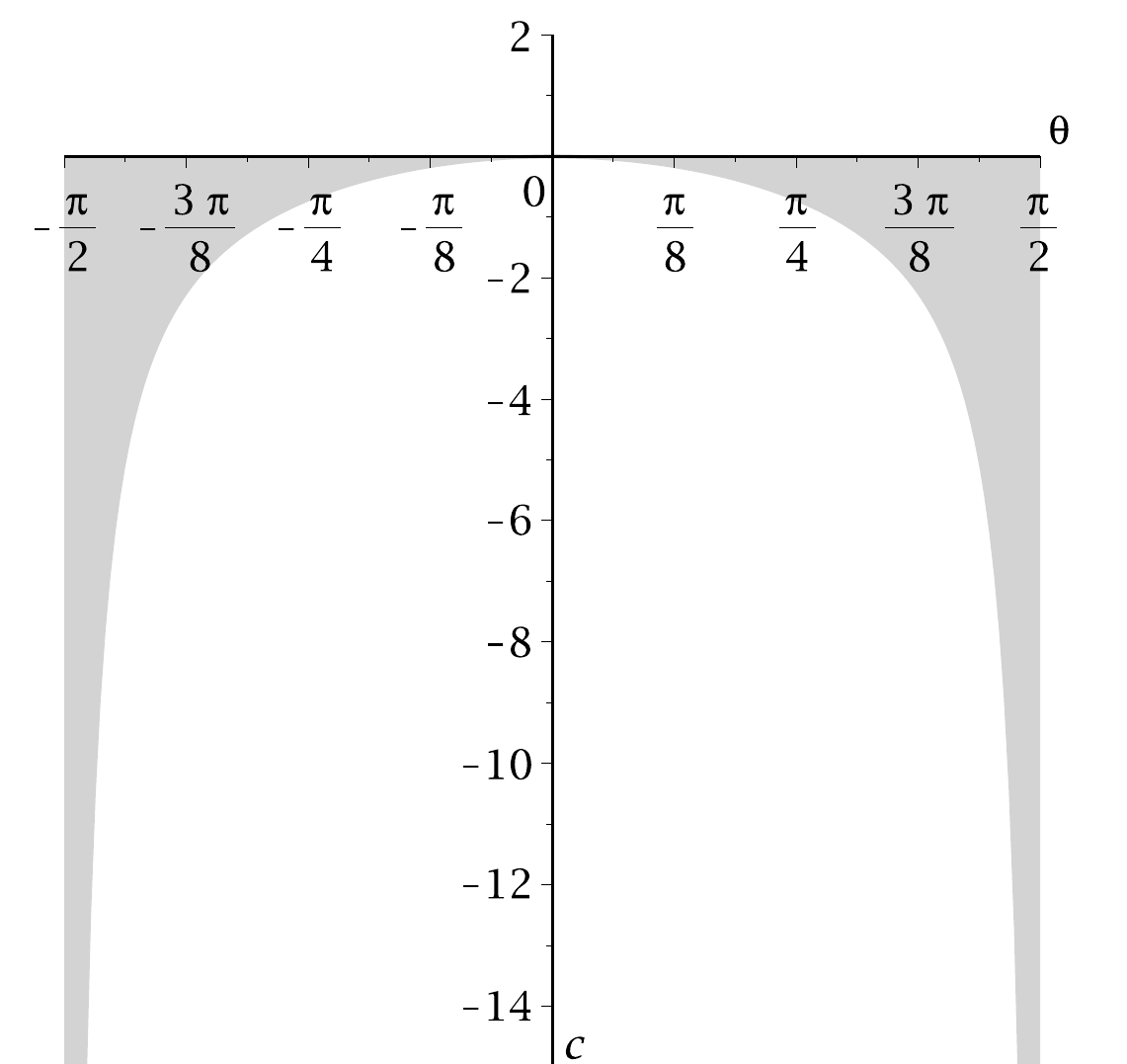}
\caption{Kinematically allowed region in $(c,\theta)$ for line-solitons
in the case of defocussing with sign-changing dispersion. 
$\k^2=1$.  
}
\label{fig:sigma1isneg1-sigma2isneg1-kinregion-ksqis1}
\end{figure}

\begin{figure}[h]
\centering
\includegraphics[width=0.45\textwidth,trim=2cm 15cm 8cm 2cm,clip]{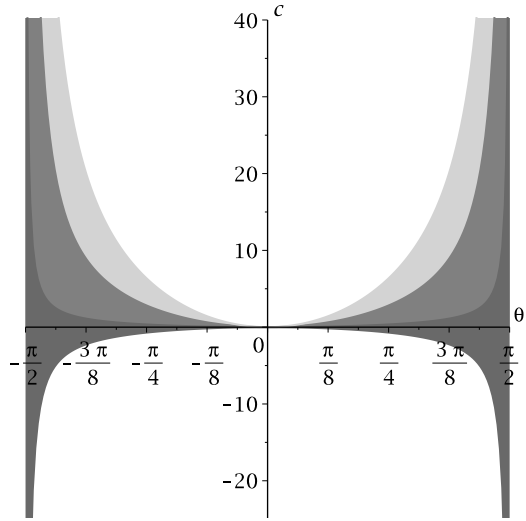}
\caption{Kinematically allowed region in $(c,\theta)$ for line-solitons
in the case of defocussing with sign-changing dispersion. 
$\k^2=$ 
$\tfrac{3}{2}$ (dark grey), 
$5$ (grey), 
$10$ (light grey);
lighter regions overlap with all darker regions. 
}
\label{fig:sigma1isneg1-sigma2isneg1-kinregion-ksqgrtrthan1}
\end{figure}

Hence, for a fixed speed $c_{\text{min}}<c<c_{\text{max}}$,
when $\k<1$ the angular range is 
$\vartheta(|c|)<|\theta|<\vartheta(|c|/(1-\k^2))$,
where $\vartheta$ is the angle \eqref{theta_c_rel}. 
Instead when $\k\geq1$ the angular range is 
$\vartheta(|c|)\leq |\theta| <\tfrac{1}{2}\pi$ if $c\leq 0$
and $\vartheta(c/(\k^2-1))<|\theta|<\tfrac{1}{2}\pi$ if $c>0$. 
These different kinematically allowed regions in $(c,\theta)$ are plotted 
in Figs.~\ref{fig:sigma1isneg1-sigma2isneg1-kinregion-ksqlessthan1}, 
~\ref{fig:sigma1isneg1-sigma2isneg1-kinregion-ksqis1},
and~\ref{fig:sigma1isneg1-sigma2isneg1-kinregion-ksqgrtrthan1}. 

The modified gKP line-solitons for $\theta\neq 0$ can be expressed as
\begin{equation}\label{u-solitary-defocusnegdisper}
u=\dfrac{s((q+1)(2q+1))^\frac{1}{2q}((1+c/|c_{\text{min}}|)\tan\theta)^\frac{1}{q}}
{\Big(|\k| + \sqrt{(\k^2-1)(1-c/c_{\text{max}})}\cosh\big(q\sqrt{(1+c/|c_{\text{min}}|)}\tan\theta\;(x+\tan\theta\;y-c\sec\theta\;t)\big) \Big)^\frac{1}{q}} 
\end{equation}
where the sign $s$ is given in Table~\ref{table:defocus-signs}. 
The width has the same expression \eqref{w-focusnegdisper} as in the previous case,
but the height is given by
\begin{equation}\label{h-defocusnegdisper}
h=\big((q+1)(2q+1)\big)^\frac{1}{2q}\Big( \big|\sqrt{(\k^2-1)(1-c/c_{\text{max}})}-|k|\big| \tan|\theta| \Big)^\frac{1}{q} .
\end{equation}
At a fixed speed, 
both the width and the height are symmetrical under $\theta\to -\theta$. 

At a fixed direction angle $\theta$,
the width decreases and the height increases
as the speed $c$ increases to $c_{\text{max}}$
and as the nonlinearity power $q$ increases.
The limiting cases of the minimum speed and maximum speed
are similar to what occurs for defocussing with normal dispersion.

\subsection{Kinematics of line-shocks}

Similarly to Proposition~\ref{prop:soliton-kinematic-conds},
we have the following kinematic properties
for all of the line-shock solutions. 

\begin{prop}\label{prop:shock-kinematic-conds}
The speed $c$ and the direction angle $\theta$ of all line-shock solutions
of the modified gKP equation \eqref{mgKP-scaled}
obey the kinematic condition 
\begin{equation}\label{speed-angle-Deltais0}
c = (\sigma_2+ \k^2) \sin^2 \theta/\cos\theta ,
\quad
\sigma_1=-1 . 
\end{equation}
No further conditions are required by the single bright/dark line-shock \eqref{bd-shock}. 
The bright line-shock \eqref{b-shock} 
and the symmetrical pair of bright/dark line-shocks \eqref{bdsymm-shockpair}
each require the additional kinematic condition \eqref{speed-angle-Cispos}. 
\end{prop}

In particular,
line-shock solutions exist only in the defocussing case, $\sigma_1=-1$.
These solutions can be expressed as
\begin{equation}\label{u-shock}
u=\dfrac{\tilde s((q+1)(2q+1))^\frac{1}{2q}(s\k \tan\theta)^\frac{1}{q}}
{\Big( 1+\exp\big(q|\k|\tan|\theta|\;(x+\tan\theta\;y-(\sigma_2+ \k^2) \tan^2 \theta \;t)\big) \Big)^\frac{1}{q}}
\end{equation}
where $s$ and $\tilde s$ are signs for the 3 types of line-shocks
shown in Table~\ref{table:shock-signs}.

\begin{table}[h!]
\caption{Signs in the line-shock}
\label{table:shock-signs}
\centering
\begin{tabular}{c|c||c|c|c}
\hline
Type
& $q$
& $s$
& $\tilde s$
& $\sgn\,\theta$
\\
\hline
\hline
bright/dark symmetrical pair
&
even
&
$1$
&
$\pm 1$
&
$\sgn\,\k$
\\
\hline
bright/dark 
&
odd
&
$1$ 
&
$1$ 
&
$\gtrless 0$
\\
\hline
bright
&
half-integer
&
$1$
&
$1$
&
$\sgn\,\k$
\\
\hline
\end{tabular}
\end{table}

We will divide the subsequent discussion into the two cases: 
defocussing with normal and sign-changing dispersion, $\sigma_2 =+1,-1$.

\subsubsection{Defocussing with normal dispersion}

When $\sigma_2=1$,
the speed is given by 
\begin{equation}\label{shock-c-normaldisper}
c(\theta) = \big(1 + \k^2\big)\sin^2\theta/\cos\theta \geq 0,
\end{equation}  
which is non-negative and increases with $\k$. 
As a consequence, for a fixed speed $c>0$, 
the direction angle is 
$\theta=\vartheta(c/(1+\k^2))$, 
where $\vartheta$ is the angle \eqref{theta_c_rel}.

\begin{figure}[h]
\centering
\includegraphics[width=0.45\textwidth,trim=2cm 15cm 8cm 2cm,clip]{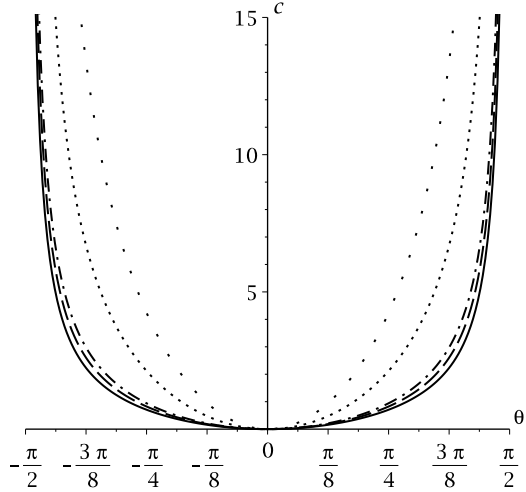}
\caption{Kinematically allowed curve in $(c,\theta)$ for line-shocks
in the case of defocussing with normal dispersion. 
$\k^2=$
$0$ (solid),
$\tfrac{1}{6}$ (dashed), 
$\tfrac{1}{3}$ (dot-dashes), 
$2$ (dots), 
$10$ (dot-spaces). }
\label{fig:shock-sigma1isneg1-sigma2is1-kinregion}
\end{figure}

The kinematically allowed curve in $(c,\theta)$ is plotted 
in Fig.~\ref{fig:shock-sigma1isneg1-sigma2is1-kinregion}. 
Note that the mKP case is given by $\k=\tfrac{1}{\sqrt{3}}$. 

The width and height of the line-shocks are given by 
\begin{align}
&\begin{aligned}  
w & = \frac{1}{q|\k|\tan|\theta|} 
= \tfrac{\sqrt{2}(\k^2+1)}{q|\k|}\big/{\textstyle \sqrt{c(\sqrt{c^2+4(k^2+1)^2}+c)}} ,
\end{aligned}
\label{w-shock-sigma2is1}
\\
&\begin{aligned}
h & =\big((q+1)(2q+1)\big)^\frac{1}{2q}\big(|\k|\tan|\theta|\big)^\frac{1}{q}
\\
& = \Big( \tfrac{(q+1)(2q+1)\k^2}{2(\k^2+1)} c(\sqrt{c^2+4(\k^2+1)^2}+c) \Big)^\frac{1}{2q} .
\end{aligned}
\label{h-shock-sigma2is1}
\end{align}
Both expressions are symmetrical under $\theta\to -\theta$. 

As the speed $c$ increases and as the nonlinearity power $q$ increases,
the width decreases and the height increases. 
In the limiting case of the minimum speed $c=0$,
the line-shock flattens to $u=0$. 

\subsubsection{Defocussing with sign-changing dispersion}

\begin{figure}[h]
\centering
\includegraphics[width=0.45\textwidth,trim=2cm 15cm 8cm 2cm,clip]{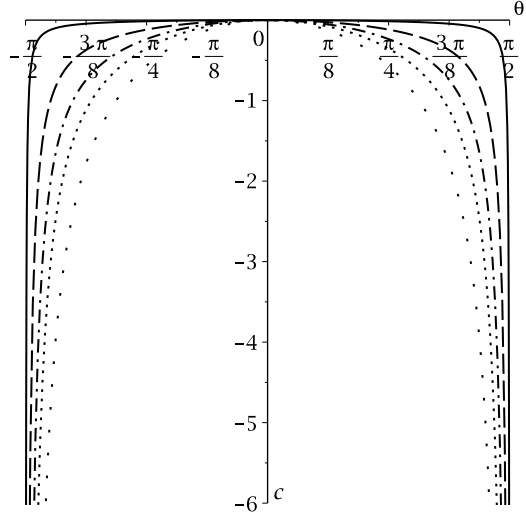}
\caption{Kinematically allowed curve in $(c,\theta)$ for line-shocks
in the case of defocussing with sign-changing dispersion. 
$\k^2=$ 
$\tfrac{1}{5}$ (dot-spaces), 
$\tfrac{1}{2}$ (dots), 
$\tfrac{2}{3}$ (dot-dashes), 
$\tfrac{5}{6}$ (dashes),
$0.98$ (solid). 
}
\label{fig:shock-sigma1isneg1-sigma2isneg1-kinregion-ksqlessthan1}
\end{figure}

\begin{figure}[h]
\centering
\includegraphics[width=0.45\textwidth,trim=2cm 15cm 8cm 2cm,clip]{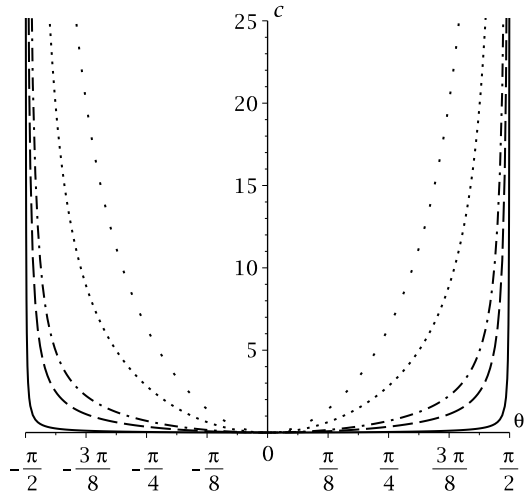}
\caption{Kinematically allowed curve in $(c,\theta)$ for line-shocks
in the case of defocussing with sign-changing dispersion. 
$\k^2=$
$1.05$ (solid),
$\tfrac{3}{2}$ (dashes),
$2$ (dot-dashes), 
$5$ (dots), 
$10$ (dot-spaces). }
\label{fig:shock-sigma1isneg1-sigma2isneg1-kinregion-ksqgrtrthan1}
\end{figure}

When $\sigma_2=-1$,
the speed is given by 
\begin{equation}\label{shock-c-signnegdisper}
c(\theta) = \big(\k^2-1\big)\sin^2\theta/\cos\theta , 
\end{equation}
which has the sign $\sgn\,c(\theta) = \sgn(|k|-1)$.
The different kinematically allowed curves in $(c,\theta)$
for $|k|<1$ and $|k|>1$ are plotted 
in Figs.~\ref{fig:shock-sigma1isneg1-sigma2isneg1-kinregion-ksqlessthan1} 
and~\ref{fig:shock-sigma1isneg1-sigma2isneg1-kinregion-ksqgrtrthan1}.

Surprisingly, for $k^2=1$,
the speed vanishes at all direction angles,
and thus the line-shock is stationary. 

For all values of $k$,
the width and height of the line-shocks are given by 
\begin{align}
&\begin{aligned}  
w & = \frac{1}{q|\k|\tan|\theta|} 
= \tfrac{\sqrt{2}|\k^2-1|}{q|\k|}\big/{\textstyle \sqrt{|c|(\sqrt{c^2+4(k^2-1)^2}+|c|)}} ,
\end{aligned}
\label{w-shock-sigma2isneg1}
\\
&\begin{aligned}
h & =\big((q+1)(2q+1)\big)^\frac{1}{2q}\big(|\k|\tan|\theta|\big)^\frac{1}{q}
\\
& = \Big( \tfrac{(q+1)(2q+1)\k^2}{2|\k^2-1|} |c|(\sqrt{c^2+4(\k^2-1)^2}+|c|) \Big)^\frac{1}{2q} .
\end{aligned}
\label{h-shock-sigma2isneg1}
\end{align}
Both expressions are symmetrical under $\theta\to -\theta$ and $c\to -c$. 

As the absolute speed $|c|$ increases and as the nonlinearity power $q$ increases,
the width decreases and the height increases. 
In the limiting case of the minimum speed $c=0$,
the line-shock flattens to $u=0$.

\section{Concluding remarks}\label{sec:remarks}

We have studied several fundamental aspects of
a recently derived general modified KP-like equation \eqref{genmKP}
and its $p$-power generalization \eqref{mgKP} 
which we refer to as the \emph{modified gKP equation}.
The general modified KP-like equation arises as the governing equation
for phase modulations of travelling waves in a universal nonlinear system in 2+1 dimensions. 
As a consequence, it is expected to model general nonlinear wave phenomena
exhibiting cubic nonlinearity, dispersion, and small transversality in 2+1 dimensions. 
Its $p$-power generalization is natural to consider from the viewpoint of analysis
and will have physical applications in modelling wave phenomena that are characterized by higher nonlinearity.

Compared to the gKP equation \eqref{gKP},
which is a $p$-power generalization of the KP equation, 
the modified gKP equation contains two extra terms that essentially involve
the $y$-derivative of the wave amplitude $u$.
Interestingly, this modified equation possesses a local variational structure
only when the coefficients of these two terms satisfy a certain relation, 
in contrast to the situation for the gKP equation. 

Our main results focus on conservation laws and line-soliton solutions.

We have derived all low-order conservation laws,
including ones that are admitted only for special powers $p$.
An interesting result is that,
apart from the variational case where Noether's theorem yields
a conserved energy and conserved momenta, 
in the non-variational case an energy is admitted only
for the modified KP-like equation \eqref{genmKP}
and not for its $p$-power generalization \eqref{mgKP}. 
Another interesting result is that in both variational and non-variational cases
these equations possess spatial flux conservation laws that yield
topological charges given by line integrals on arbitrary closed curves in the $(x,y)$-plane.
We show that these charges give rise to integral constraints on initial data for the Cauchy problem. 
This sets the stage for investigating well-posedness and related questions such as 
global (long-time) existence of solutions. 

We also have derived all line-soliton solutions
and compared them to the gKP line-solitons.
Due to the two extra terms in the modified gKP equation \eqref{mgKP},
we find that non-symmetrical bright/dark pairs of line-solitons
are supported when $p$ is even,
whereas only symmetrical bright/dark pairs arise for the gKP equation.
Moreover, 
the kinematically allowed region in the parameter space of speed and angular direction
is significantly different for the modified gKP equation \eqref{mgKP}
and has qualitatively distinct features in the cases of
focussing/defocussing, normal/sign-changing dispersion. 
In particular, for both types of dispersion in the defocussing case, 
the kinematically allowed region has an essential dependence
on the coefficients of the two extra terms. 

At the boundary of the kinematically allowed regions in the defocussing case, 
we also find that a line-shock solution is supported. 
Such solutions do not exist for the gKP equation and thus are new phenomena
produced by the extra terms in the modified gKP equation \eqref{mgKP}.

For future work, it will be interesting to investigate
the stability of the line-soliton and line-shock solutions 
and to determine how their stability may depend on 
the nonlinearity power $p$ and the size of the non-KP terms in the equation. 
It will also be interesting to study the well-posedness of the Cauchy problem
and determine the conditions under which long-time solutions exist.

\section*{Acknowledgements}
S.C.A.\ is supported by an NSERC research grant
and thanks the University of C\'adiz for additional support during the period 
when this work was initiated.

\section*{Appendix}

Here we summarize computational aspects underlying 
the classification of conservation laws for the modified gKP equation 
in Proposition~\ref{prop:multrs} and Theorem~\ref{thm:conslaws}.

The determining equation \eqref{Q-deteqn} for multipliers \eqref{low-order-Q} 
with differential order less than four 
has a splitting with respect to the set of variables $\{\partial^4 w,\partial^5 w,\partial^6 w\}$.
Note this set excludes the leading derivative $w_{xxxx}$ and its differential consequences.
The choice of this leading derivative rather than the other two possibilities $w_{tx}$ and $w_{yy}$ 
allows us to find multipliers that would otherwise appear at a higher differential order,
as explained in a general context in \Ref{Wol}. 

We have carried out the setting up and splitting of the determining equation \eqref{Q-deteqn} 
by using Maple.
This yields an overdetermined system consisting of 3356 equations 
to be solved for $Q$ as well as for $k\neq0$ and $q\neq0$,
with $\sigma_1^2=\sigma_2^2=1$. 
Solving the system is a nonlinear problem because 
$Q$ appears linearly in products with $k$ while $q$ appears nonlinearly. 
We use the Maple package `rifsimp` to find the complete case tree of solutions. 
For each solution case in the tree, we solve the system of equations 
by using Maple 'pdsolve' and 'dsolve', 
and we check that the solution has the correct number of free constants/functions
and satisfies the original overdetermined system. 
Finally, we merge overlapping cases by following the method explained in \Ref{RecAnc2017}. 
This yields the classification of multipliers listed in Proposition~\ref{prop:multrs}. 

For each of the multipliers, 
we derive the corresponding conserved density $T$ and spatial flux $(X,Y)$ 
by applying the repeated integration process \cite{Wol,BCA-book,Anc-review}
to the righthand side of the characteristic equation for each multiplier. 
This method has the advantage that we can obtain $T$, up to equivalence, 
so that it has the lowest possible differential order. 
We do all integrations with respect to spatial derivatives of $w$ first, 
whereby the remaining integrations with respect to $w_t$ will always yield terms of minimal differential order in $T$. 
This yields the form for $T$, $X$, $Y$ shown in Theorem~\ref{thm:conslaws}.

\end{document}